\documentclass{elsart}
\usepackage{amssymb}
\usepackage[fleqn]{amsmath}
\usepackage{graphicx}
\usepackage{bigstrut}
\newcommand{\ed}{\mathrm{e}}
\newcommand{\id}{\mathrm{i}}
\newcommand{\vvec}{\mathbf{v}}
\newcommand{\nvec}{\mathbf{n}}
\newcommand{\evec}{\mathbf{e}}
\newcommand{\Uvec}{\mathbf{U}}

\newcommand{\dd}{\mathrm{d}}
\newcommand{\Kd}{\mathrm{K}}
\newcommand{\Ld}{\mathrm{L}}
\newcommand{\Id}{\mathrm{I}}
\newcommand{\Dd}{\mathrm{D}}
\newcommand{\Wd}{\mathrm{W}}

\newcommand{\euler} {_{\mathrm{Euler}}}

\newcommand{\loc}   {^{\mathrm{loc}}}

\renewcommand{\crit}  {_{\mathrm{c}}}
\newcommand{\eff}  {_{\mathrm{eff}}}
\newcommand{\new}  {_{\mathrm{m}}}
\newcommand{\old}  {}
\newcommand{\Dset}{\mathbb{D}}

\newcommand{\mach}{\mathcal{M}}
\newcommand{\cheb}{Chebychev}

\renewcommand{\leq}{\leqslant}
\renewcommand{\geq}{\geqslant}

\DeclareMathOperator{\sech}{sech}

\begin{document}
\begin{frontmatter}
\title{Boundary layers and emitted excitations in nonlinear Schr\"odinger superflow past a disk}
\author[ENS,CTPmail]{Chi-Tuong Pham}
\author[LIMSI,CNmail]{Caroline Nore}
\author[ENS,MEBmail]{Marc-\'Etienne Brachet}

\address[ENS]{Laboratoire de Physique Statistique de l'Ecole Normale 
Sup{\'e}rieure, \\
associ{\'e} au CNRS et aux Universit{\'e}s Paris VI et VII,
24 Rue Lhomond, 75231 Paris, France}
\address[LIMSI]{LIMSI-CNRS, BP 133, 91403 Orsay C\'edex France,
Universit\'e Paris XI, D\'epartement de Physique, 91405
Orsay Cedex France}
\thanks[CTPmail]{pham@lps.ens.fr (correponding author)}
\thanks[CNmail]{nore@limsi.fr}
\thanks[MEBmail]{brachet@lps.ens.fr}
%
%
%
%
\begin{abstract}

The stability and dynamics of nonlinear Schr\"odinger superflows 
past a two-dimensional disk are investigated using a specially 
adapted pseudo-spectral method based on mapped \cheb\ polynomials. 
This efficient numerical method allows
the imposition of both Dirichlet and Neumann boundary conditions at the
disk border. Small coherence length boundary-layer approximations to stationary solutions 
are obtained analytically. 
Newton branch-following is used to compute the complete bifurcation diagram of 
stationary solutions. 
The dependence of the critical Mach number on the coherence length is characterized.
Above the critical Mach number, at coherence length larger than fifteen times the diameter of the disk, rarefaction pulses are dynamically nucleated,
replacing the vortices that are nucleated at small coherence length. 

\end{abstract}

\end{frontmatter}


%
%
%
%
%
\section{Introduction}

It is well known that, above a critical speed,  
superfluidity breaks down and dissipation sets in~\cite{Don91}. 
Much work has been devoted to the understanding of this phenomenon
within the mathematical description of superfluidity provided by
the nonlinear Schr{\"o}dinger equation (NLSE) also called
the Gross-Pitaevskii  equation  \cite{G61,Pit61,Lanflueng}.
 
The NLSE can be used to describe two quite different physical systems: 
superfluid $^4$He and Bose-Einstein condensates of ultra-cold atomic vapor.

In the case of superfluid $^4$He, the NLSE can be considered as a valid mathematical model
provided that the temperature is low enough for the normal fluid to be negligible.
This is clearly the case in recent experiments~\cite{NSHC03} that are 
performed at temperatures below $130$ mK. 
Note that the excitations of superfluid 
$^4$He are accurately described by the famous Landau
spectrum which includes phonons in the low wave number range, and maxons
and rotons in the high (atomic-scale) wave number range. 
In contrast, the standard NLSE 
(the equation used in the present article)
only contains phonon excitations. It therefore incompletely represents
the atomic-scale excitations in superfluid $^4$He.
However, there exist straightforward generalizations of the NLSE
\cite{PR93b,RB01} that do reproduce the correct excitation spectrum,
at the cost of introducing a spatially non-local interaction potential.
For reasons of simplicity we shall not use such generalizations in the present article.

Since Bose-Einstein condensation in dilute gases in traps was
experimentally observed~\cite{AEMWC95,DMADDKK95,BSTH95},
this field is in rapid evolution: recent results include
the production and detection of an isolated quantized vortex~\cite{MAHHWC99,IGRCGGLPK01},
the nucleation of several vortices~\cite{MCWD00} and details
of vortex dynamics~\cite{RBD02}.
The dynamics of these compressible nonlinear quantum fluids is accurately
described by the NLSE allowing direct
quantitative  comparison between theory and experiment \cite{DGPS99}.

The stability of Bose-Einstein condensates (BEC) in the presence of a moving obstacle can thus be studied in the framework of the NLSE.
Raman {\it et al.} have studied dissipation
in a Bose-Einstein condensed gas by moving a blue detuned
laser beam through the condensate at different velocities \cite{RKODKHK99}.
In their inhomogeneous condensate, they observed
a critical Mach number for the onset of dissipation that was compared with the NLSE predictions.
 
In their pioneer work, Frisch, Pomeau and Rica~\cite{FPR92} 
performed direct numerical simulations of the NLSE 
to study the stability of two-dimensional superflows around a disk. 
They observed a transition
to a dissipative regime characterized by vortex nucleation that
they interpreted in terms of a saddle-node bifurcation of
the stationary solutions of the NLSE. Later, using numerical branch-following
techniques, Huepe and Brachet~\cite{HB97,HB00} obtained the complete
bifurcation diagram in which the stable and unstable branches
are connected through a saddle-node bifurcation.
Asymmetric solutions were also found, generated by a secondary 
pitchfork bifurcation of the stable branch.
The symmetric and asymmetric unstable solutions correspond respectively
to two and one vortices. 
The critical speed was shown to converge, for small coherence length,
to the Eulerian value computed by Rica~\cite{R01}.  Three-dimensional effects
leading to a lowering of the critical speed were also considered~\cite{NHB00}.

In all the above numerical studies, the effect of the two-dimensional disk was 
represented in the NLSE by a simple repulsive potential. 
Thus no boundary conditions were applied and
the numerical results were (weakly) dependent on the details of the repulsive potential.

One of the main motivations of the present paper is to obtain numerical results that are reliable
(i.e. do not depend on an {\it ad hoc} artificial repulsive potential)
at finite value of the coherence length. 
We will thus consider the NLSE as a partial differential equation
with standard boundary conditions applied on the disk.
This mathematical problem will be studied by using an efficient
pseudo-spectral method, based on angular Fourier series and radially mapped \cheb\ polynomials,
that was specifically designed for the present study. 
The numerical solutions will be compared with analytic boundary layer approximations, 
that are valid for small velocity and coherence length.
Similar expansions were performed for a spherical obstacle in~\cite{BR00}.

The paper is organized as follows: section~\ref{sec:calcul} contains the governing equations;
section~\ref{sec:bl} is devoted to the derivation of the boundary layer analytical expressions
for Dirichlet conditions; 
in section~\ref{sec:new_method}, we describe the new specially designed pseudo-spectral 
method;  section~\ref{sec:stat} contains validations of the numerical procedure and
new results on bifurcation diagrams and critical Mach numbers; 
in section~\ref{sec:dyn}, our results on the dynamically emitted excitations are reported,
with emphasizing on the nucleation of rarefaction pulses; 
finally  section~\ref{sec:conclusion} is our conclusion.
More details on the numerical method are found in the appendix 
where the resolutions needed to obtain spectral convergence are discussed.

%
%
\section{Governing equations}\label{sec:Definitions}

\label{sec:calcul}

In this section, we present the hydrodynamic form of the NLSE that models 
the effect of a disk of radius unity (diameter $D=2$), moving at constant 
speed $\vvec=v\evec_x$ in a two-dimensional superfluid at rest. In the frame 
of the disk, the system is equivalent to a superflow around a disk, with constant 
speed $-\vvec$ at infinity. Let $\Omega$ be the plane $\Cset$ deprived of $\Dset$ 
the disk of radius unity and $\partial \Omega$ the boundary of the domain, that is the circle of radius unity. We will naturally use the polar coordinates $(r,\theta)$
such that $x=r\cos\theta$ and $y=r\sin\theta$ and the associated unit vectors are
denoted by $ (\evec_r, \evec_\theta)$.
The system can then be described with the following action functional
\begin{equation}
\mathcal{A} [\psi,\bar{\psi}] = \int \dd t \left\{\sqrt{2} c \xi \int_\Omega \dd^2 x \frac \id 2 \left[\bar{\psi}\partial_t\psi - {\psi}\partial_t\bar{\psi}\right] - \mathcal{F}_0 \right\} \label{eq:NLS_func}
\end{equation}
where $\psi$ is a complex field, $\bar\psi$ its conjugate. The speed of sound $c$ and the
 so-called healing length $\xi$ are the physical parameters of the system. $\mathcal{F}_0$ is the energy of the system that reads
\begin{align}
&\mathcal{F}_0 [\psi,\bar{\psi}] = \mathcal{E}  - \vvec\cdot \mathcal{P}
\label{eq:F_nls_2D} 
\end{align}
with
\begin{align}
&\mathcal{E} [\psi,\bar{\psi}] = c^2 \int_\Omega \dd^2 x \left[\xi^2\vert\nabla\psi \vert^2 + {\frac 1 2}  (\vert\psi\vert^2-1)^2\right]\label{eq:E_nls_2D}\\
&\mathcal{P} [\psi,\bar{\psi}]= \sqrt{2}c\xi\int_\Omega\dd^2 x  \frac \id 2 \left[\vphantom{\frac 1 2}(\psi-1) \nabla \bar{\psi} - (\bar{\psi}-1)\nabla\psi\right]. \label{eq:P_nls_2D}
\end{align}
The presence of the constants $-1$ in Eq. (\ref{eq:P_nls_2D}) ensures the convergence of the integral. The Euler-Lagrange equation corresponding to (\ref{eq:NLS_func}) provides the NLSE
\begin{equation}
\id\partial_t\psi= \frac{c}{\sqrt{2}\xi}\left[-\xi^2\Delta\psi-\psi+\vert\psi\vert^2\psi\right] +\id\vvec\cdot \nabla\psi\,,\label{eq:nls2D_ESNL}
\end{equation}
defined in the domain $\Omega$. 
This equation can be mapped into two hydrodynamical equations by applying Madelung's transformation \cite{Don91}
\begin{equation}
\psi = \sqrt{\rho} \exp\left(\frac{\id \phi}{\sqrt{2}c\xi}\right),
\end{equation}
that defines a fluid of density $\rho$ and velocity
\begin{equation}
\Uvec = \nabla \phi - \vvec
\end{equation}
The real and imaginary parts of the NLSE yield the following equations of motion
\begin{align}
&\partial_t\rho  + \nabla\cdot (\rho\Uvec) = 0\label{eq:eqcontinuite}\\
&\partial_t\phi = -\frac 1 2 (\nabla\phi)^2 + c^2 (1-\rho) + c^2\xi^2 \frac{\Delta \sqrt{\rho}}{\sqrt{\rho}} + \vvec\cdot\nabla\phi.\label{eq:eqBernoulli}
\end{align}
These equations correspond respectively to the continuity and the Bernoulli equations 
(with a supplementary quantum pressure term) for a barotropic compressible and irrotational flow. 
Note that two non-dimensional parameters control the system: the Mach number $\mach=|\vvec|/c$
(where $\vvec$ is the flow velocity at infinity and $c$ the sound speed) and the ratio of the
healing length $\xi$ to the diameter of the disk $D$. 
In the limit $\xi/D \rightarrow 0 $, the quantum pressure term vanishes and we recover 
the system of equations describing an Eulerian flow. 
We now investigate the problem of the boundary conditions on the obstacle.


In previous studies~\cite{HB00}, boundary conditions were applied by adding to the NLSE a repulsive potential term 
strong enough to force the density to zero inside the disk. 
In the present work, we consider the mathematically standard Dirichlet and Neumann boundary conditions
that will both be directly imposed at the border of the obstacle.
 
\subsection{Dirichlet boundary conditions}

The Dirichlet boundary conditions read $\psi|_{r=1}=0$.
They thus prescribe zero density on the obstacle and correspond 
to the presence of an unpenetrable obstacle (a laser in a BEC or a solid obstacle in superfluid $^4$He). 
They correspond to the following conditions, in hydrodynamical variables: 
first, the condition on $\rho$ is obviously
\begin{equation}
\rho=0 \qquad \text{at $r=1$}
\end{equation}

Second, the square root of the density $R=\sqrt\rho$ being constant on the obstacle, we have $\partial_t R|_{r=1}=0$ and $\partial_\theta R|_{r=1}=0$. The continuity equation (\ref{eq:eqcontinuite}) expressed in term of $R$ then yields $\partial_r R \cdot U_\perp|_{r=1} =0$, so that the Dirichlet conditions also imply
\begin{equation}
U_\perp = \partial_r \phi -v\cos\theta= 0\qquad \text{ at }r=1 
\end{equation}

\subsection{Neumann boundary conditions}

The Neumann boundary, in hydrodynamical variables, read
\begin{align}
&\partial_r\rho=0 &&\text{at $r=1$}\label{eq:NeumannBC_rho}\\
&U_\perp = \partial_r \phi -v\cos\theta= 0&&\text{at $r=1$} \label{eq:NeumannBC_phi}
\end{align}
They correspond to the following  conditions in term of the complex field $\psi$: $\partial_r(\psi \exp{\left(\frac{\id  v r_0^2 \cos \theta}{\sqrt{2}c\xi r}\right)})|_{r=r_0=1}=0$. 
Note that, compared to the Dirichlet conditions, the Neumann conditions are more academic than physically realistic. 
Nevertheless, it is interesting to study the influence of such boundary conditions on the stationary solutions of the problem, 
especially their effects on the boundary layer on the obstacle. 
For instance, one could think, that with such conditions the stationary solution would be ``closer'' to that of the Eulerian flow than with Dirichlet conditions. 
We will see below that the situation is more complex. 
A more physical motivation to study such boundary conditions is related to the problem of capillary-gravity free surface flows past a cylindrical obstacle, where the quantum pressure term is replaced by a capillary term. In this related problem, the Neumann conditions are physical ones~\cite{bib:these_CTP}.

%
%
%
%
%
%
%
\section{Boundary layer solutions -- Analytical results}
\label{sec:bl}
We now present calculations of the stationary solutions in the limit $\xi/D\rightarrow 0$. For non-zero Mach number,
\begin{align}
&\mach = |\vvec|/c
\end{align}
we define the new phase variable~\cite{R01}
\begin{align}
&\varphi = -(\phi-vr\cos\theta)/v.
\end{align}
The Bernoulli (\ref{eq:eqBernoulli}) and continuity (\ref{eq:eqcontinuite}) equations then read
\begin{align}
 & 0=\xi^2\frac {\Delta \sqrt{\rho}}{\sqrt{\rho}} -\rho + 1+ \frac {\mach^2}{2}[1-(\nabla\varphi)^2]\label{eq:2D_Bernoulli}\\
& 0=\rho \Delta\varphi + \nabla\rho\cdot\nabla\varphi.\label{eq:2D_continuite}
\end{align}
The  Dirichlet boundary conditions now read
\begin{align*}
&\rho|_{\partial\Omega} = 0\\
&\partial_r\varphi|_{\partial\Omega} = 0.
\end{align*}
At finite but small Mach number, we expand $\rho$ and $\varphi$ as 
\begin{align}
&\rho = \rho^{\langle 0\rangle} + \mach^2\rho^{\langle 1\rangle}+ \cdots +\mach^{2k}\rho^{\langle k\rangle}+ \cdots\\
&\varphi = \varphi^{\langle 0\rangle} + \mach^2\varphi^{\langle 1\rangle}+ \cdots +\mach^{2k}\varphi^{\langle k\rangle}+ \cdots.
\end{align}
Note that if one knows $\varphi$ at order $\mach^{2k}$, on can formally deduce  $\rho$ at order $\mach^{2(k+1)}$ by solving (\ref{eq:2D_Bernoulli}). The potential  $\varphi$ can then be computed at order $\mach^{2(k+1)}$ by 
solving (\ref{eq:2D_continuite}). 
In order to compute $\varphi$, we will have to solve equations of the type
\begin{align}
\frac{\dd^2 y}{\dd r^2}(r) + \frac 1 r \frac{\dd y}{\dd r}(r) - \frac {1}{r^2} y(r) = \mathrm{RHS}(r)
\end{align}
Solutions to the corresponding homogeneous equation are
\begin{align}
y(r) = Ar + Br^{-1}
\end{align}
so that the general equation with nonzero righthandside $\mathrm{RHS}(r)$ can be computed using the method of variation of parameter. Using the boundary conditions $\lim_{r\rightarrow +\infty} y(r) = 0$ and  $\dd y/\dd r (r=1)=0$ yields for the solution of the inhomogeneous equation the explicit expression 
\begin{multline}
y(r)= -\frac 1 {2r}  \int_1^{+\infty} \mathrm{RHS}(u)(1+u^2)\dd u  \\ - \frac r 2 \int_r^{+\infty}\mathrm{RHS}(u)\dd u + \frac 1 {2r} \int_r^{+\infty} u^2 \mathrm{RHS}(u) \dd u\label{eq:var_cstes}
\end{multline}
provided that the function $\mathrm{RHS}$ decreases rapidly enough at infinity. Note that the first term of $y(r)$ yields a term of the type $C/r$. Due to the expressions of $\mathrm{RHS}$ encountered  in the following computations, the two last terms will turn out to tend to zero exponentially (on a length scale of order $\xi$), so that the behavior at infinity of the function $y$ will be governed by a long-range  algebraic term that reads
\begin{equation}
y(r)\underset{r\rightarrow +\infty}{\sim} -\frac 1 {2r}  \int_1^{+\infty} \mathrm{RHS}(u)(1+u^2)\dd u \label{eq:equivalent_var_cstes}
\end{equation}

We now turn to the computation of  the stationary Dirichlet solution. Expressions for $\rho^{\langle 0\rangle} $ and $\varphi^{\langle 0\rangle}$ are obviously needed to bootstrap the iteration. They are obtained by the following considerations.

When  the Mach number is zero, $\varphi=0$ is solution of the stationary equations and $\rho$ satisfies 
\begin{equation}
\xi^2\frac {\Delta \sqrt{\rho}}{\sqrt{\rho}} -\rho + 1 =0 \label{eq:R}
\end{equation}
Writing $\rho(r,\theta)=R^2(r)$ yields the equation 
\begin{equation}
\xi^2\Delta R +R-R^3 = \xi^2 (\partial_{rr} + \frac 1 r \partial_r)R+R-R^3 = 0
\end{equation}
with boundary conditions $R(1)=0$. 
A first approximation for the solution of this equation, obtained by neglecting the term $(\xi^2/r)\partial_r R$,  reads
\begin{equation}
R^{\langle 0\rangle}_0 = \tanh\left(\frac{r-1}{\sqrt 2 \xi}\right)  \label{eq:R0machnul}
\end{equation}
This result, valid up to order $\xi$, can be improved by setting $R=R^{\langle 0\rangle}_0+R^{\langle 0\rangle}_1$. 
Inserting $R$ in (\ref{eq:R}), collecting the terms of order $\xi$ and solving the resulting differential equation yields, after tedious computations,
\begin{equation}
R^{\langle 0\rangle} _1=\frac{\xi}{6{\sqrt{2}}}\left[{-3 - \cosh 2s + 
    \left( 4 + 3s \right) \sech^2s + 
    \sinh 2s + 3\,\tanh s}\right] \label{eq:R1machnul}
\end{equation}
where $s
=(r-1)/\sqrt 2 \xi$.
Thus the explicit expression
\begin{equation}
\rho^{\langle 0\rangle} =\rho^{\langle 0\rangle}_0+\rho^{\langle 0\rangle}_1=( R_0^{\langle 0\rangle}) ^2 + 2 R_0^{\langle 0\rangle}   R^{\langle 0\rangle} _1 \label{eq:rhomachnul}
\end{equation}
gives the correct approximation to the density, up to order $\xi^2$
 in the limit $\xi/D\rightarrow 0$.

The velocity potential $\varphi^{\langle 0\rangle}$ satisfies
\begin{equation}
\Delta \varphi^{\langle 0\rangle} = -\nabla \rho^{\langle 0\rangle}\cdot\nabla \varphi^{\langle 0\rangle}+(1-\rho^{\langle 0\rangle})\Delta \varphi^{\langle 0\rangle}.  \label{eq:phi0_mach_non_nul_rhop0.ne.0}
\end{equation}
We write $\varphi^{\langle 0\rangle}=\varphi\euler^{\langle 0\rangle}+\tilde\varphi^{\langle 0\rangle}$ where $\varphi\euler^{\langle 0\rangle}=(r+1/r)\cos\theta$ is the solution at order $0$ in $\mach^2$ of the Eulerian flow. 
Using the relation $\Delta \varphi\euler^{\langle 0\rangle}=0$, equation (\ref{eq:phi0_mach_non_nul_rhop0.ne.0}) yields the
following equation for $\tilde\varphi^{\langle 0\rangle}$
\begin{equation}
\Delta \tilde\varphi^{\langle 0\rangle} = 
- \nabla\rho^{\langle 0\rangle}\cdot\nabla\varphi\euler^{\langle 0\rangle} 
- \nabla\rho^{\langle 0\rangle}\cdot\nabla\tilde\varphi^{\langle 0\rangle}+(1- \rho^{\langle 0\rangle})\Delta \tilde \varphi^{\langle 0\rangle} \label{eq:phi0_mach_non_nul_rhop0.ne.0.bis}
\end{equation}
This equation cannot be solved directly. We thus proceed to a perturbative development by writing $ \tilde\varphi^{\langle 0\rangle}= \tilde\varphi^{\langle 0\rangle}_1+ \tilde\varphi^{\langle 0\rangle}_2$ where $ \tilde\varphi^{\langle 0\rangle}_1$ is of order $\xi$ and $ \tilde\varphi^{\langle 0\rangle}_2$ of order $\xi^2$. 
In the right hand side of equation (\ref{eq:phi0_mach_non_nul_rhop0.ne.0.bis}), one can keep at the dominant order of our computations the first term and drop the two others. The function $ \tilde\varphi^{\langle 0\rangle}_1$ is then solution of the equation
\begin{equation}
\Delta \tilde\varphi^{\langle 0\rangle}_1 =-\nabla \rho^{\langle 0\rangle}_0\cdot\nabla \varphi\euler^{\langle 0\rangle} \label{eq:phi0_mach_non_nul_rhop0.ne.0.approx00}
\end{equation}
The expression of  $\tilde\varphi^{\langle 0\rangle}_1$ can be computed using Eq. (\ref{eq:var_cstes}). Eq. (\ref{eq:equivalent_var_cstes}) yields~\cite{PNB03} 
\begin{equation}
\tilde\varphi^{\langle 0\rangle}_1 \underset{r\rightarrow +\infty}{\sim} \frac{2\sqrt{2}\xi-4(\log 2) \xi^2}{r} \cos\theta
\end{equation}

In order to obtain the full correction at order $\xi^2$ of the $1/r$-algebraic term we also need to compute $\tilde\varphi^{\langle 0\rangle}_2$ which verifies
\begin{equation}
\Delta \tilde\varphi^{\langle 0\rangle} _2= 
- \nabla \rho^{\langle 0\rangle}_1\cdot\nabla \varphi\euler^{\langle 0\rangle}   
- \nabla\rho^{\langle 0\rangle}_0\cdot\nabla\tilde\varphi^{\langle 0\rangle}_1+(1-\rho^{\langle 0\rangle}_0)\Delta \tilde \varphi^{\langle 0\rangle}_1
\end{equation}
Using again Eq. (\ref{eq:equivalent_var_cstes}), a lengthy computation yields
\begin{equation}
\tilde\varphi^{\langle 0\rangle}_2 \underset{r\rightarrow +\infty}{\sim} \xi^2 \frac{10-4\log 2}{3r}  \cos\theta
\end{equation}
The velocity potential $\varphi^{\langle 0\rangle}$ thus reads
\begin{equation}
\varphi^{\langle 0\rangle} = \left[r 
+ \left(1+2\sqrt 2 \xi+\frac{10-16\log 2}{3}\xi^2 + O(\xi^3)\right) \frac 1 r + \varphi^{\langle 0\rangle}_{\mathrm{loc}}(r)\right] \cos\theta
\end{equation}
where $\varphi^{\langle 0\rangle}_{\mathrm{loc}}$  exponentially vanishes  at infinity.

Note that the compressible Eulerian flow around a disk of radius $r_1$ admits at order zero in $\mach^2$ the following solution 
\begin{equation}
\varphi^{\langle 0\rangle}_{\mathrm{Euler}, r_1} = \left(r+ \frac{r_1^2}{r}\right)\cos\theta \label{eq:Euler_general}
\end{equation}
in order to satisfy the boundary condition $\partial_r\varphi|_{r=r_1}=0$. Thus, the correction to $\varphi^{\langle 0\rangle}_{\mathrm{Euler}}$ is a long-range term that can be physically interpreted as a renormalization of the diameter of the disk : at large distances the superflow is equivalent to an Eulerian flow around a disk of radius $r\eff$ given by
\begin{equation}
\left(\frac{r\eff}{r_0}\right)^2 = 1 + 2\sqrt{2}\left(\frac{\xi}{r_0}\right)+\frac{10-16 \log 2}{3} {\left(\frac{\xi}{r_0}\right)}^2 + O(\xi^3). \label{eq:rayon_effectif_dir}
\end{equation}

The order $\xi$ term was first computed in \cite{PNB03}. 
Similar results were obtained directly, using matched
expansions, for a spherical obstacle in~\cite{BR00}. 
This reference also includes the governing matched expansion equations for the case of a 2D disk, however the authors did not solve these equations.

The same procedure with Neumann boundary conditions can be shown to lead to a renormalized radius~\cite{bib:these_CTP}
\begin{equation}
\left(\frac{r\eff}{r_0}\right)^2 = 1 - \frac{3}{2}\mach^2 \left(\frac{\xi}{r_0}\right)^2
\label{reff_neumann}
\end{equation}
Note that contrary to the case
of Dirichlet conditions, this effective size is dependent on the Mach number, which was not the case
for Dirichlet conditions. It is also smaller than the corresponding Dirichlet effective value.

%
%

\section{Specially adapted pseudo-spectral method} \label{sec:new_method}

We have specifically developed a code that can
accurately accommodate both large-$r$ asymptotic behavior and thin boundary layers near 
the obstacle at $r=1$. It is based on a \cheb\ decomposition using an adequate mapping.
It allows us to consider a unique obstacle in contrast with 
periodic pseudo-spectral methods~\cite{HB00} which in fact model a network of obstacles.

\subsection{Mapping for a unique obstacle}

Using standard polar coordinates $\{\theta,r\}$, together with the relation
\begin{equation}
r(z)=z^{-1}
\label{simplemap}
\end{equation}
the domain $\{0\leq \theta< 2 \pi,-1\leq z\leq 1\}$, can be mapped 
into the physical domain
$\{x,y\}$, with $x^2+y^2\geq 1$.

The basic mapping is
\begin{align}
&x=z^{-1} \cos{\theta}\qquad\qquad y=z^{-1}  \sin{\theta}
\label{polar}
\end{align}
and the inverse transformation reads
\begin{align}
&z= \pm \frac 1 {\sqrt{x^2+y^2}}\qquad\qquad\theta= \arctan(y/x) +\frac {\pi \mp \pi} 2
\label{polarinv}
\end{align}

Any generic real field $\Psi(x,y)$ ($\Psi$ stands for 
$\mathrm{Re}\phi$,  $\mathrm{Im}\psi$, etc.)  appearing in the encountered
equations of motion
is expressed in the $\{\theta,z\}$ domain as  
\begin{equation}
\Psi(\theta,z)=\Psi(x(\theta,z),y(\theta,z))
\end{equation}
with
$x(\theta,z)$ and $y(\theta,z)$ defined in (\ref{polar}).

As $x(\theta,z)=x(\theta+\pi,-z)$ and $y(\theta,z)=y(\theta+\pi,-z)$, 
the $\{x,y\}$ domain is mapped twice unto the $\{\theta,z\}$ domain.
A mapped field must therefore satisfy
\begin{equation}
\Psi(\theta,z)=\Psi(\theta+\pi,-z)
\label{symmetry}
\end{equation}

The equations of motion are expressed
as partial differential equations in the $\{\theta,z\}$ domain by writing the
differential operators $\nabla$ and $\Delta$ in terms of 
$\theta$ and $z$ derivatives that are polynomial in $z$, e.g. ${\Delta \psi}  =  z^2\frac{\partial^2 \psi}{\partial \theta^2} + z^4\frac{\partial^2 \psi}{\partial z^2}  + z^3\frac{\partial \psi}{\partial z}$.

\subsection{Generalization of the mapping}

In the special case where $\xi/D$ is large, we found useful to generalize the $r(z)=1/z$ mapping to
\begin{equation}
r_\lambda(z_\lambda)=\frac\lambda {z_\lambda} + (1-\lambda) z_\lambda
\label{mapgeneralise}
\end{equation}
This mapping has the same overall characteristics than $1/z$ and reduces to it for $\lambda=1$. For  $\lambda>1$ it stretches the coordinate, thereby increasing the resolution at large distance by moving the collocation points away from the $r=1$ disk. Generalizations to expressions (\ref{polar}) and (\ref{polarinv}) 
 are easily derived. 

\subsection{Spatial discretization} \label{Spatial Discretization}

The field $\psi$ is  spatially discretized, in the $(\theta,z)$ domain,  using a standard \cheb-Fourier
pseudo-spectral method \cite{GO77}, based on the expansion
\begin{equation}
\psi(\theta,z)=
\sum_ {n=1-{N_\theta / 2}}^{N_\theta / 2} \left\{
\sum_ {p=0}^{N_r} 
\psi_{n,p}
  \,T_{p}(z)\right\}\,\exp{\id n \theta}
\label{specrep}
\end{equation}
where $T_p(z)=\cos{(p\arccos{z})}$ is the order-$p$ \cheb\ polynomial and $N_\theta$ and $N_r$ represent resolutions.

The pseudo-spectral method calls for using  fast Fourier 
transforms to evaluate (\ref{specrep}) on the collocation points grid $(\theta_m,z_k)$ 
with
\begin{align}
&\theta_m=\frac{2 \pi m}{N_\theta} ; && 0\leq m<N_\theta&&\\
&z_k=\cos{\frac{\pi k}{N_r}} ; && 0\leq k \leq N_r&&
\label{colloc_grid}
\end{align}
The relation $T_n(\cos{x})=\cos{n x}$ reduces
the \cheb\ transform appearing in 
(\ref{specrep}) to a (fast) Fourier cosine transform. Thus, the evaluation of 
(\ref{specrep}) (and its inverse) only requires a time proportional to 
$N_\theta N_r \log (N_\theta N_r)$. Computations of nonlinear terms are carried 
out on the grid representations, while $\theta$ and $z$ derivatives are carried out
on the \cheb -Fourier representations.

The main virtue of mapping (\ref{polar}) together with expansion (\ref{specrep})
is its ability to accurately accommodate both large-$r$ asymptotic behavior and thin boundary layers near $r=1$.
Indeed, on the one hand, (\ref{specrep}) is an expansion in product of polynomials in $r^{-1}$ with functions  $\cos{n \theta}$ and $\sin{n \theta}$, precisely the type of functions needed to capture large-$r$ behavior (see section~\ref{sec:num_bl} and \cite{R01}).
On the other hand, the accumulation of collocation points $z_p$ (see equation (\ref{colloc_grid})) and the regularity of (\ref{polar}) near $z=\pm 1$ allows expansion (\ref{specrep}) to simultaneously resolve boundary layers at $r=1$ with thickness of order $1/N_r^2$ \cite{GO77}.

\subsection{Spectral symmetries of the fields}

 As $\psi$ is real, the coefficients $\psi_{n,p}$ in
(\ref{specrep}) are complex conjugate
\begin{equation}
\psi_{-n,p}=\bar{\psi}_{n,p}
\end{equation}
They obey an additional relation, stemming from (\ref{symmetry}).
Setting $z=cos(\theta')$, the fields must be invariant under the transformation
$\theta\mapsto\theta+\pi$, $\theta'\mapsto\theta'+\pi$. In spectral space, this transformation reads
$\psi_{n,p}\mapsto(-1)^n(-1)^p\psi_{n,p}$, implying
\begin{equation}
\psi_{n,p}=(-1)^{n+p}\psi_{n,p}
\end{equation}
Thus the $\psi_{n,p}$ coefficients are non-zero only when $(n,p)$ are jointly even or jointly odd.
This relation, similar to that found in the Taylor-Green vortex \cite{NAB97}, 
is used to speed-up the evaluation of (\ref{specrep}) by a factor $2$, using specially designed even-odd Fast Fourier transforms.

Integral of mapped fields are performed on the collocation points using the discrete formula
\begin{equation}
\int_\Omega r\dd r\dd \theta \psi(r,\theta)= - \frac{2\pi}{N_\theta} \frac{\pi}{N_r}\sum_{n=0}^{N_\theta -1} \sum_{p=0}^{N_r/2 -1} \psi(\theta_m, z_k) \sqrt{1-z_p^2} \frac{\dd r}{\dd z}(z_p) r(z_p)  
\end{equation}

\section{Stationary solutions -- Numerical results} \label{sec:stat}

This section is devoted to
the numerical determination of stationary solutions using the branch-following method detailed in the appendix.
We first focus on the particular case of the Eulerian flow (that is when $\xi/D=0$). 
This case has been previously investigated using methods based on series in Mach 
number by Rica \cite{R01}, and
the critical Mach number is known with great precision. 
We next compare analytical results of section~\ref{sec:bl} with numerically obtained
profiles of boundary layers with Dirichlet conditions.
It is thus a good test of the numerical precision and efficiency 
of our new method, presented above in section \ref{sec:new_method}.

The rest of the section contains the numerical results on the bifurcation diagrams and the stationary solutions of the NLSE at small and large coherence lengths, for the two types of boundary conditions: Dirichlet and Neumann. We discuss the dependence on $\xi/D$ of the critical Mach number.

\subsection{Eulerian limit}\label{sec:euler_limit}

In the limit $\xi/D\rightarrow 0$, the NLSE turns into the equations of an Eulerian compressible flow
\begin{align}
&\partial_t\phi = -\frac 1 2 (\nabla\phi)^2 + c^2 (1-\rho)  + \vvec\cdot\nabla\phi\label{eqn:EulerBernoulli}\\
&\partial_t\rho = -\rho \Delta\phi -\nabla\rho\cdot\nabla\phi + \vvec\cdot\nabla\rho\label{eqn:Eulercontinuite}
\end{align}%
that are respectively the Bernoulli and continuity equations. We now search for their stationary solutions. 
Note that, knowing the stationary field $\phi$, the Bernoulli equation yields explicitly an expression of $\rho$ that reads
\begin{equation}
\rho=1  -\frac 1 {2c^2} (\nabla\phi)^2   + \frac 1 {c^2} \vvec\cdot\nabla\phi 
\label{eq:explicitrhoEuler}\end{equation}
Therefore, $\phi$ is solution to the following equation
\begin{equation}
0=-\rho \Delta\phi -\nabla\rho\cdot\nabla\phi + \vvec\cdot\nabla\rho
\end{equation}
with density $\rho$ given by Eq. (\ref{eq:explicitrhoEuler}) and  a unique boundary condition on the disk (instead of two in the NLSE case)
\begin{equation}
U_\perp = \partial_r \phi -v\cos\theta= 0\qquad \text{at $r$ =1}\label{bc_Euler}
\end{equation}

Using the branch following method presented in the appendix 
yields the numerical stationary solutions of the two-dimensional Eulerian flow with respect to the Mach number. The critical Mach 
number is then the one at which the local Mach number 
\begin{equation}
\mach\loc = \frac{|\Uvec|}{c\loc}= \frac{|\nabla{\phi}-\vvec|}{\sqrt{\rho}}. 
\end{equation}
reaches $1$, at $(x=0,y=\pm 1)$\cite{R01}.

The value of the computed critical Mach number determined in this way depends on the resolution. It is found to decrease when the 
resolutions in $\theta$ and $r$ increase.
In order to obtain $11$ significant digits, the (minimum) needed resolution is 
$N_\theta \times N_r = 512 \times 32$. 
The critical Mach number then found is $\mach\crit=0.36969705259(9)$. 
In order to obtain the same precision as that of series methods  \cite{R01} ($\mach\crit^{\mathrm{Rica}} = 0.36969(7)$), it is sufficient to use 
the resolution $N_\theta \times N_r = 128 \times 16$, that is only $8$ radial grid points in physical space.

\subsection{Comparison with analytical boundary layer results for Dirichlet condtions} \label{sec:num_bl}

\begin{figure}[ht!]
\centerline{\includegraphics[width=.45 \textwidth]{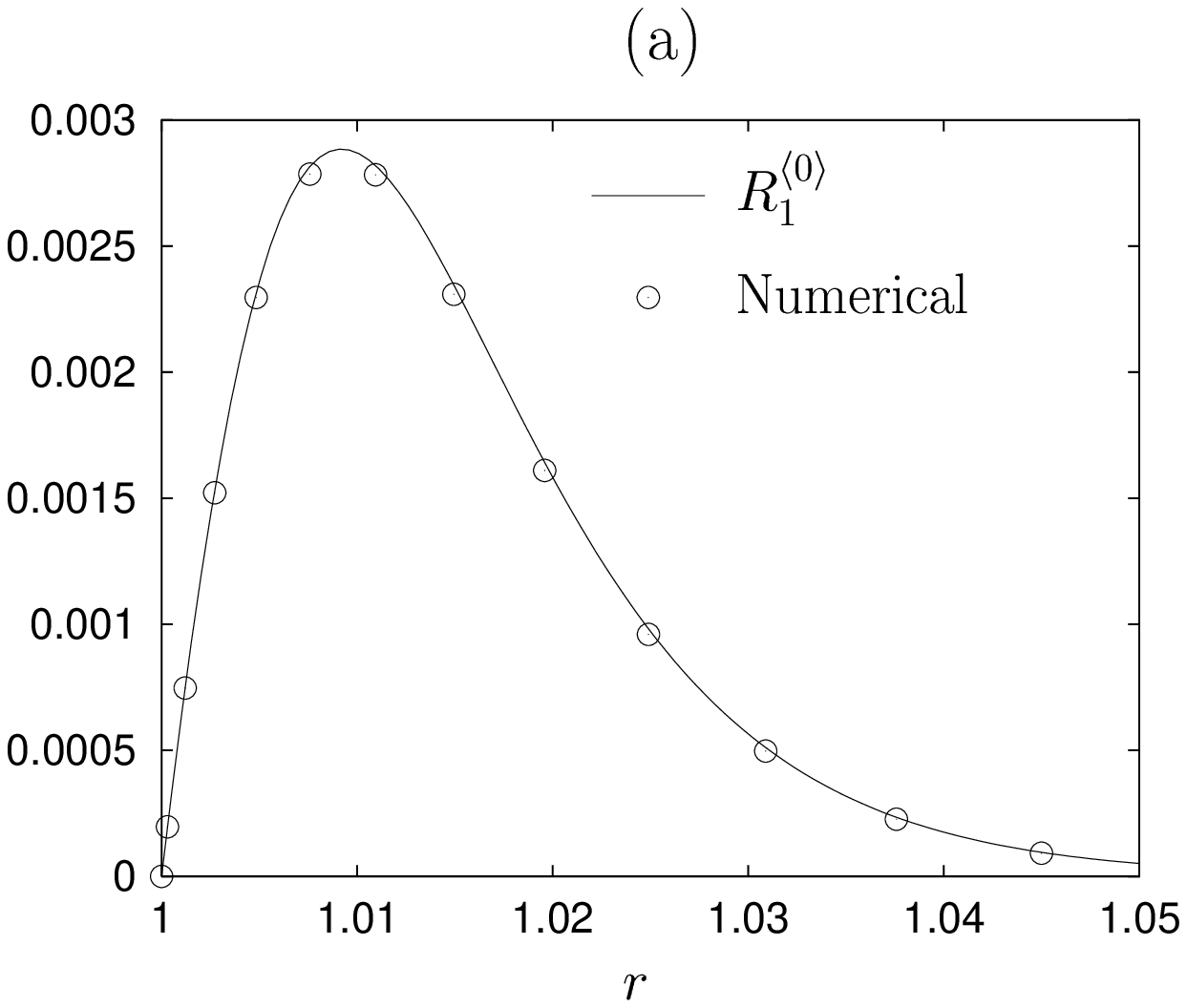}\hfill \includegraphics[width=.45 \textwidth]{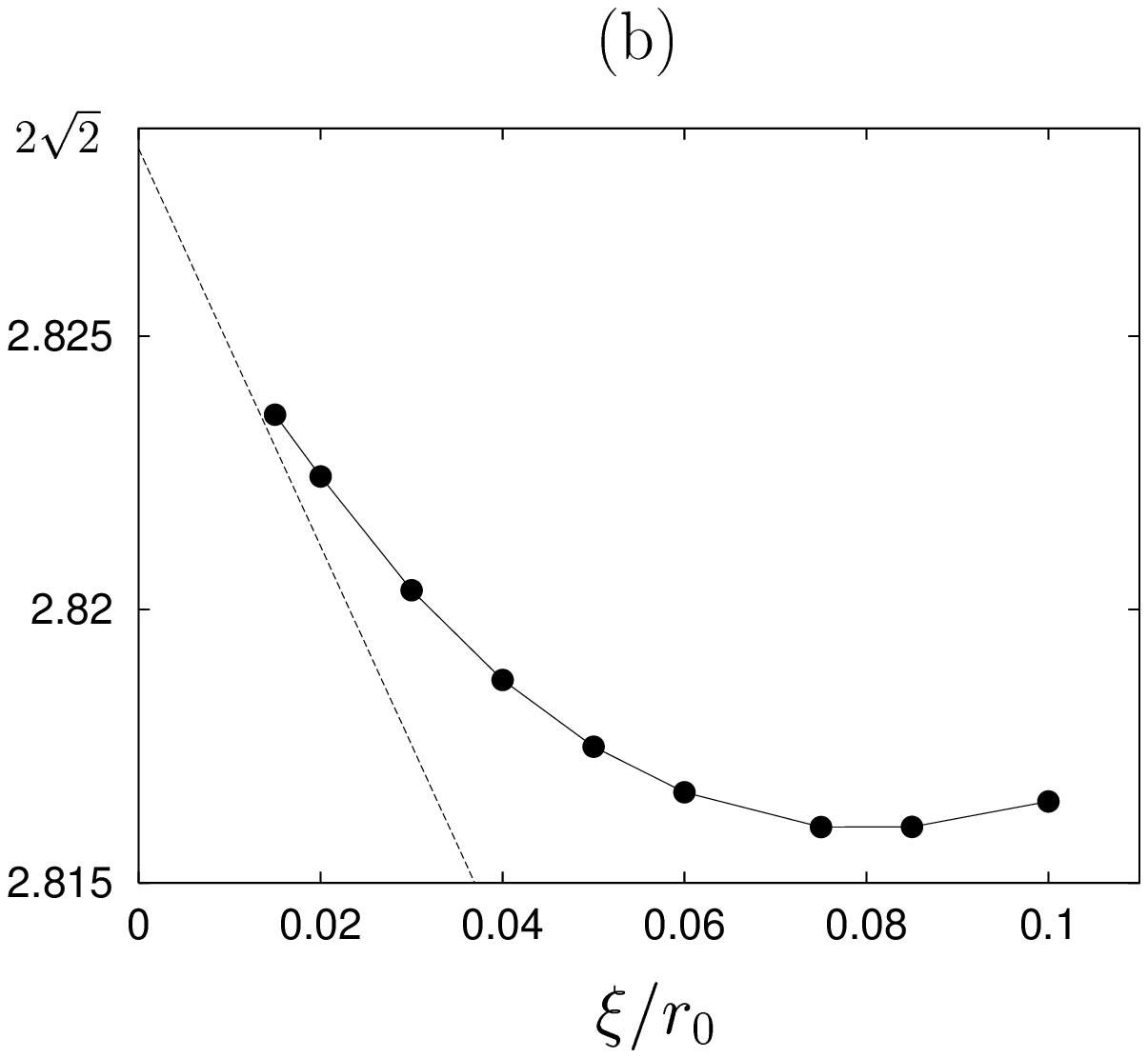}}
\caption{(a) Plot of $R^{\langle 0\rangle}_1(s(r))$ with $s(r) = (r-1)/\sqrt 2 \xi$ (see Eq. (\ref{eq:R1machnul}) and below) together with the function  $R^{\mathrm{num}}(r)-\tanh((r-1)/\sqrt 2 \xi)$ where $R^{\mathrm{num}}(r)$ is the numerical result obtained with our numerical method for $\xi/D = 1/200$ and $\mach=0$. The agreement is excellent. (b) 
Calculation of $\delta\eff$ as a function of $\xi$ together with the curve
$2\sqrt 2 + (10-16 \log 2)(\xi/r_0)/3$ (see text). The difference between the two curves is due to the term in $\delta\eff$ of higher order in $\xi$.
Note that  the agreement is very good for small $\xi/r_0$.}
\label{fig:profil_num}
\end{figure}

We now compare the analytical results of section~\ref{sec:bl} with numerically obtained
profiles of boundary layers with Dirichlet conditions. Figure~\ref{fig:profil_num}(a) displays  boundary layer profiles of the density square-root ($R=\sqrt \rho$) computed at $\xi/D= 1/200 \text{ and } \mach = 0$.

To stress the agreement between analytical and numerical results, it
is more convenient to substract the term (\ref{eq:R0machnul})
$R^{\langle 0\rangle}_0 =\tanh((r-1)/\sqrt 2 \xi)$ in the numerical profiles
and compare the higher order terms thus obtained with the analytical expression.

In the same way, Figure~\ref{fig:profil_num}(b) presents the order $\xi^2$ variation
of the effective radius as the following combination
$\delta\eff=((r\eff/r_0)^2-1)/(\xi/r_0))$. The line $2\sqrt 2 + [10-16 \log 2)/3](\xi/r_0) $ is
shown on the same graph since expression~(\ref{eq:rayon_effectif_dir}) predicts
that
\begin{equation} 
\delta\eff= 2\sqrt 2 + \frac{10-16 \log 2}{3} {\left(\frac{\xi}{r_0}\right)} + O(\xi^2).
\label{deff}
\end{equation} 
The value of $(r\eff/r_0)^2$ was extracted from the  numerical results  by calculating the coefficient in $\cos\theta/r$ of the velocity potential of the stationary state substracted by  the corresponding Eulerian flow coefficient (see Eq. (\ref{eq:Euler_general})).

The agreement between analytical and numerical results is very good
for small $\xi/D$ emphasizing the ability of our method to compute thin boundary
layers.

\subsection{Bifurcation diagrams and stationary states at small coherence length}

We present the bifurcation diagrams and the stationary solutions of the NLSE at small coherence length, 
for the two types of boundary conditions: Dirichlet and Neumann. 

The numerical methods presented in section~\ref{sec:new_method} and 
the appendix also converge very well in the NLSE case.
For instance for Dirichlet conditions, the resolution needed to compute a whole bifurcation diagram is  lower than in previous studies by Huepe and Brachet~\cite{HB00}. With the present method, the resolution needed in the case $\xi/D=1/20$ is $N_\theta\times N_r = 64 \times 64$ whereas Huepe \etal\ needed a spatial (rectangular) resolution $N_x \times N_y = 256 \times 128$ for the same ratio $\xi/D$. The gain in resolution is then of  a factor $8$. This factor increases for smaller $\xi/D$.
These excellent convergence properties are
detailed in the appendix, section \ref{sec:app_small}.

\subsubsection{Bifurcation diagrams}

\begin{figure}[ht!]
\centerline{\includegraphics[width=.45\textwidth]{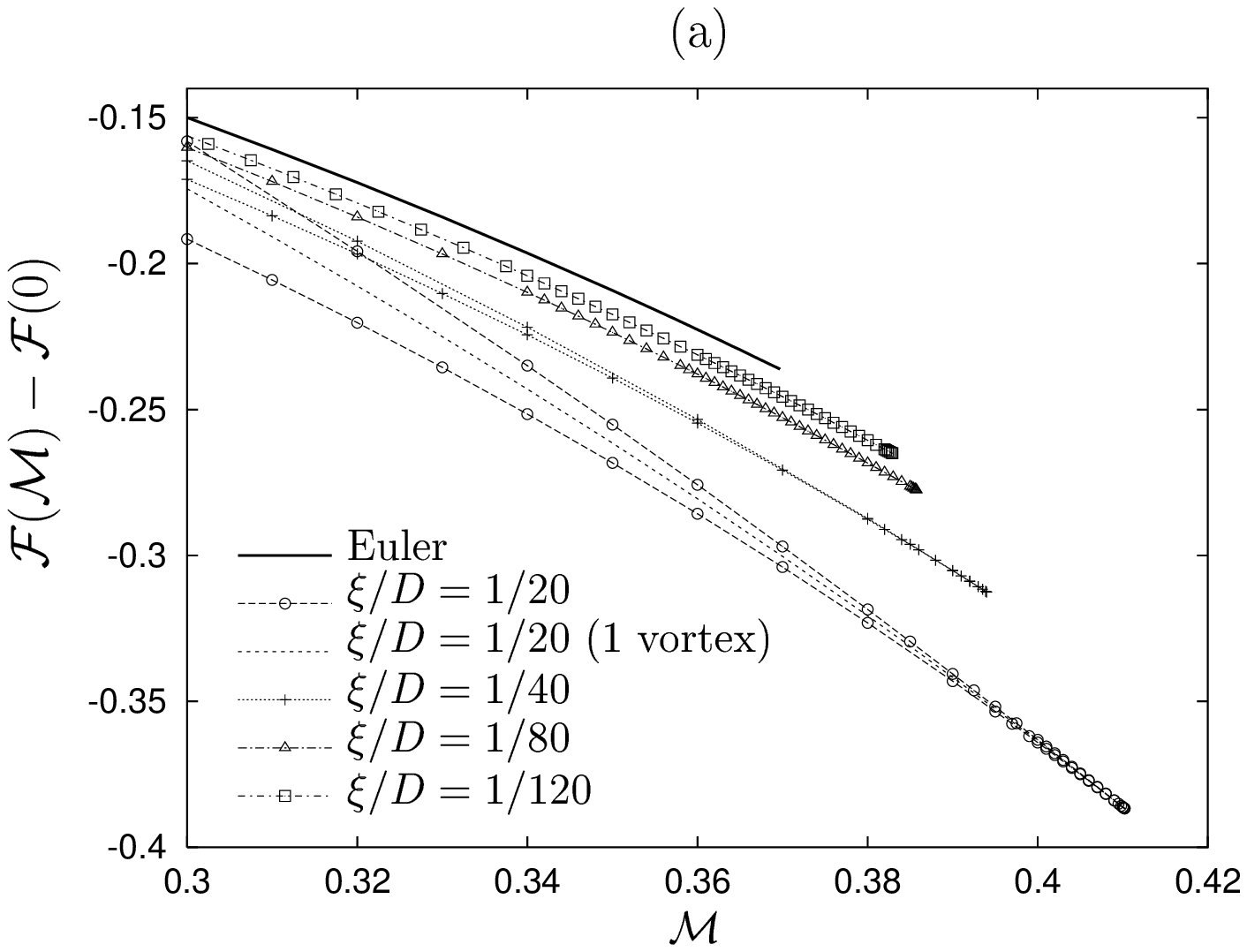}\qquad\includegraphics[width=.45\textwidth]{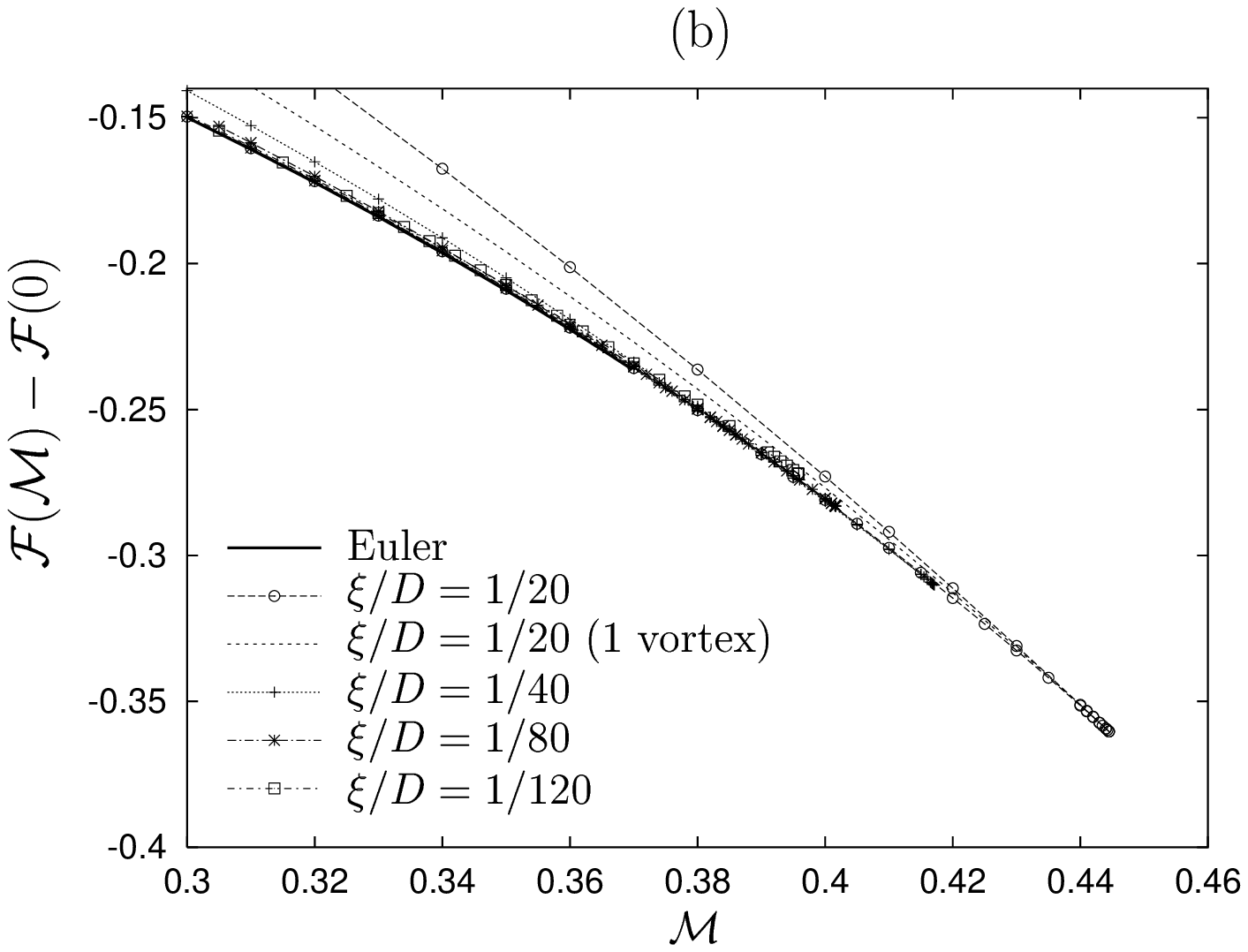}
}
\caption{Bifurcation diagrams for small coherence lengths. 
Energy functional $\mathcal{F}(\mach)-\mathcal{F}(0)$ versus Mach number. 
(a) Dirichlet conditions; (b) Neumann conditions.  For $\xi/D=1/20$,
the asymmetric unstable solution branch is represented (it stands for a one-vortex branch stemming from a pitchfork bifurcation).
At a fixed Mach number, the energy difference between a stable and an 
asymmetric unstable solution is roughly half the energy difference
between a stable and a symmetric unstable solution.}
\label{fig:nls_2D_diag_bif_small_xi}
\end{figure}

In order to study bifurcation diagrams, we define a new free energy by:
\begin{align}
&\mathcal{F} [\psi,\bar{\psi}] =\mathcal{F}_0 [\psi,\bar{\psi}] - \vvec\cdot  \sqrt{2}c\xi\oint_{\partial\Omega} \dd\ell \nvec \frac {1}{2\id}(\psi-\bar\psi),  \label{eq:Fp_nls_2D}
\end{align}  
with $\nvec = - \evec_r$ the unit vector normal to the boundary.
The rightmost term in (\ref{eq:Fp_nls_2D}) does not affect the equation of motion 
and is always zero for the $\psi|_{\partial\Omega} = 0$ Dirichlet boundary.
For Neumann conditions, this term ensures that a stationary solution $\psi_0$
is an actual extremum of the functional $\mathcal{F}$, {\it i.e.}, $\mathcal{F}$
satisfies 
$\mathcal{F}[\psi_0+\delta \psi, \bar{\psi_0}+\bar{\delta \psi}] - \mathcal{F}[\psi_0, \bar{\psi_0}]=0$ 
at first order in $\delta\psi$.
This property implies the existence of a generic cusp in $\mathcal{F}$ at the bifurcation point
(see figure~\ref{fig:nls_2D_diag_bif_small_xi}).

For simplicity, we will use the notation $\mathcal{F}(\mach)=\mathcal{F}[\psi_0(\mach), \bar{\psi_0}(\mach)]$.
The values  of $\mathcal{F}(\mach) - \mathcal{F}(0)$
(the change of energy $\mathcal{F}$, relative to zero Mach number)
is displayed in figure~\ref{fig:nls_2D_diag_bif_small_xi} 
as a function of the Mach number $\mach$
for various values of $\xi/D$ and the two types of boundary conditions.
As can be seen by inspection of the figure,
for each $\xi/D$, the stable branch (solid line) disappears with
the unstable solution (dashed line) at a
saddle-node bifurcation when $\mach = \mach\crit$.
There are no stationary solutions beyond this point.
This qualitative behavior
is the signature of a Hamiltonian saddle node bifurcation.

By inspection of figure~\ref{fig:nls_2D_diag_bif_small_xi}(b), we can see that the stable stationary branches for Neumann conditions are almost superimposed on the Euler branch which is not the case for Dirichlet conditions. This is due to the fact that Dirichlet conditions impose a zero of the density at the border of the disk, contrary to Neumann conditions and Eulerian flow.

\begin{figure}[ht!]
\centerline{\includegraphics[width=.4\textwidth, angle=-90]{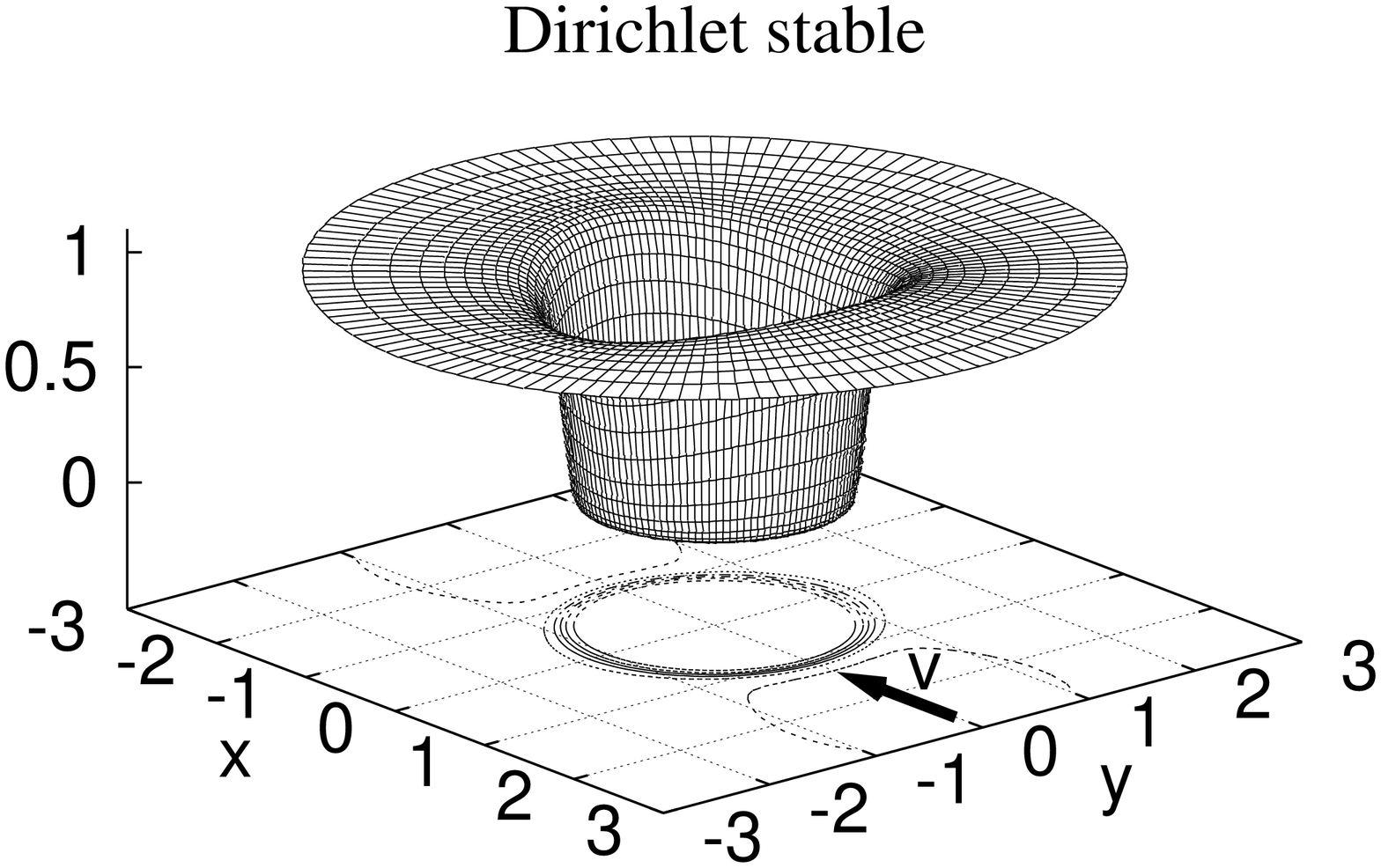}\hfill\includegraphics[width=.4\textwidth, angle=-90]{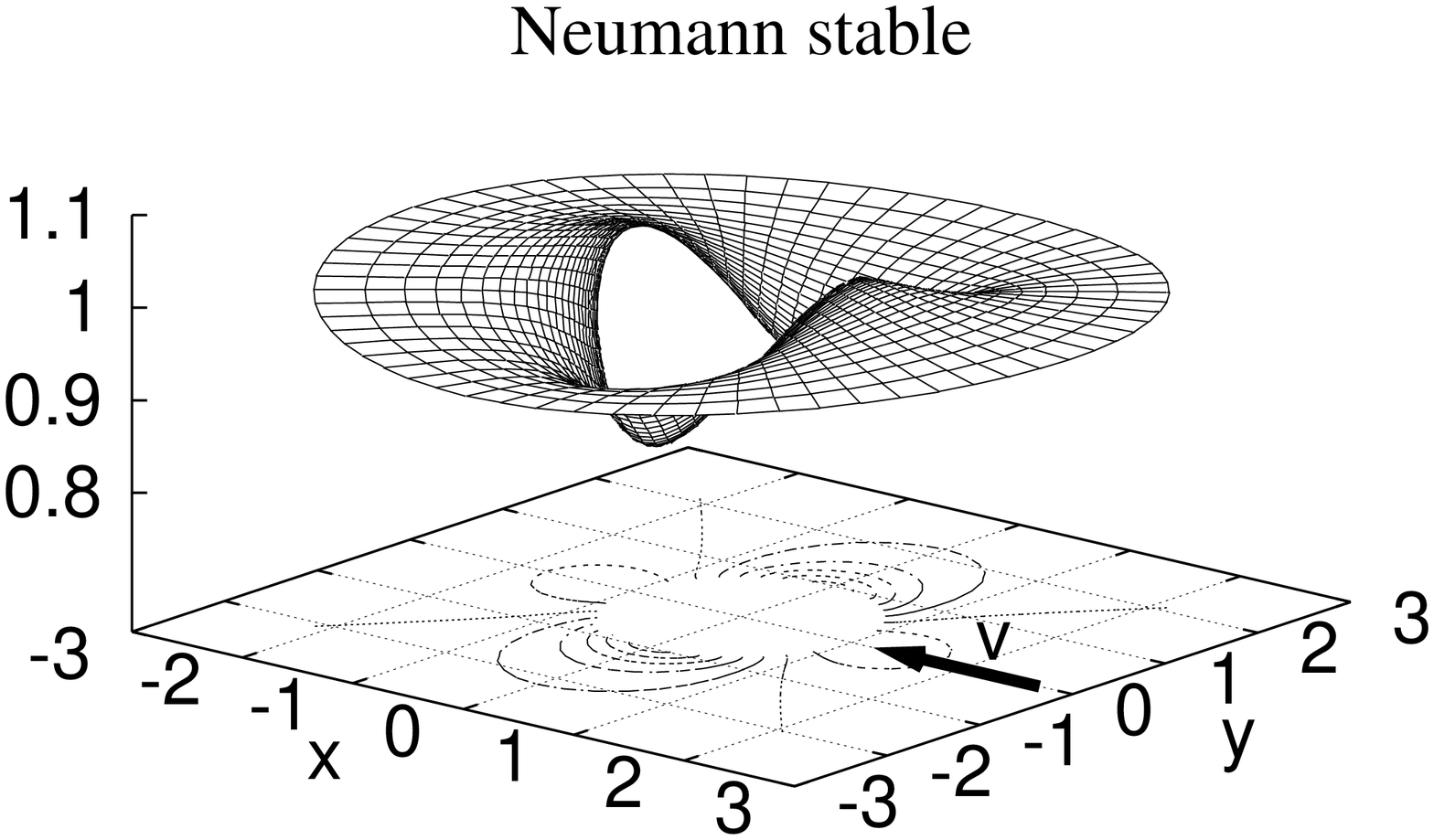}}
\centerline{\includegraphics[width=.4\textwidth, angle=-90]{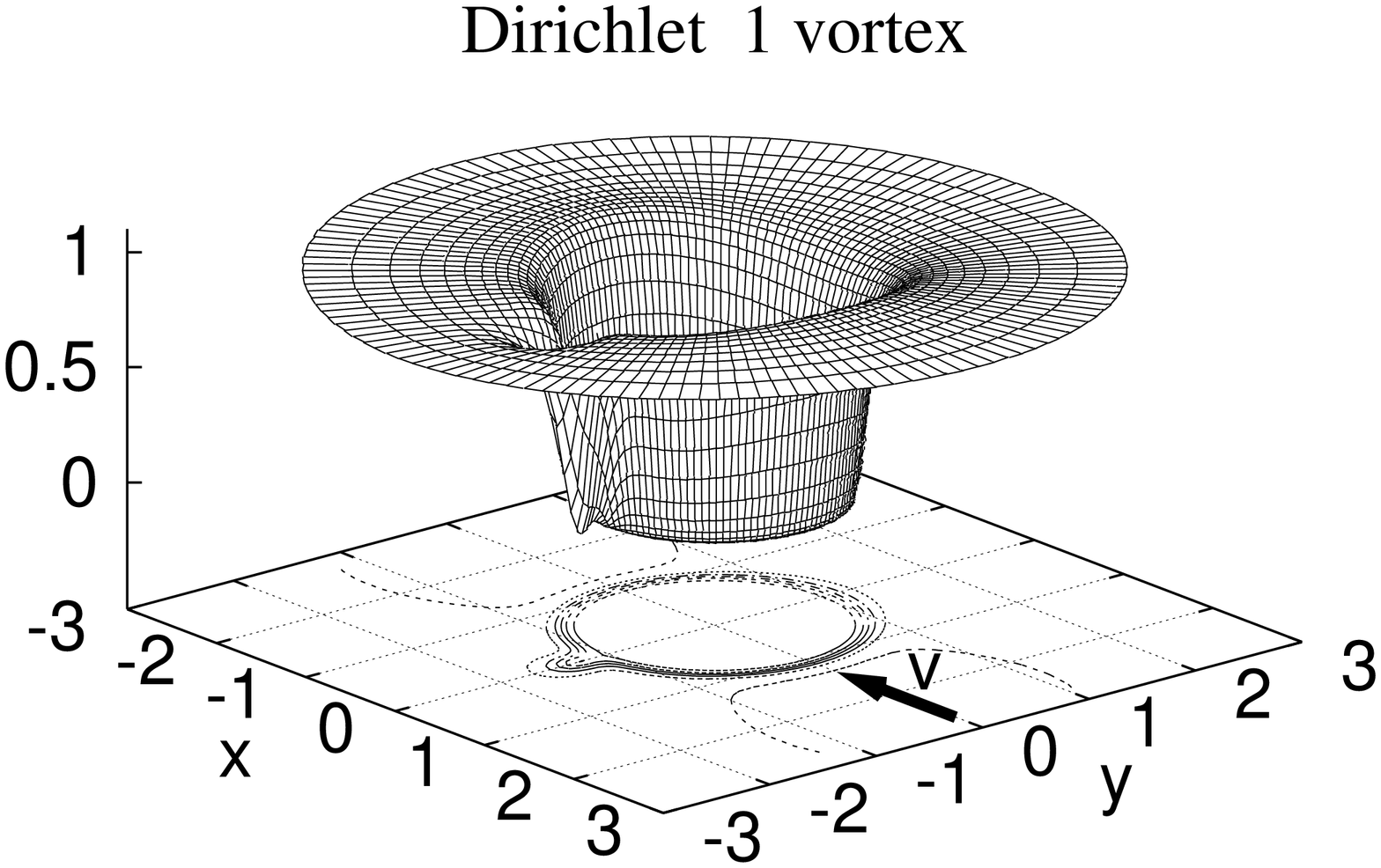}\includegraphics[width=.4\textwidth, angle=-90]{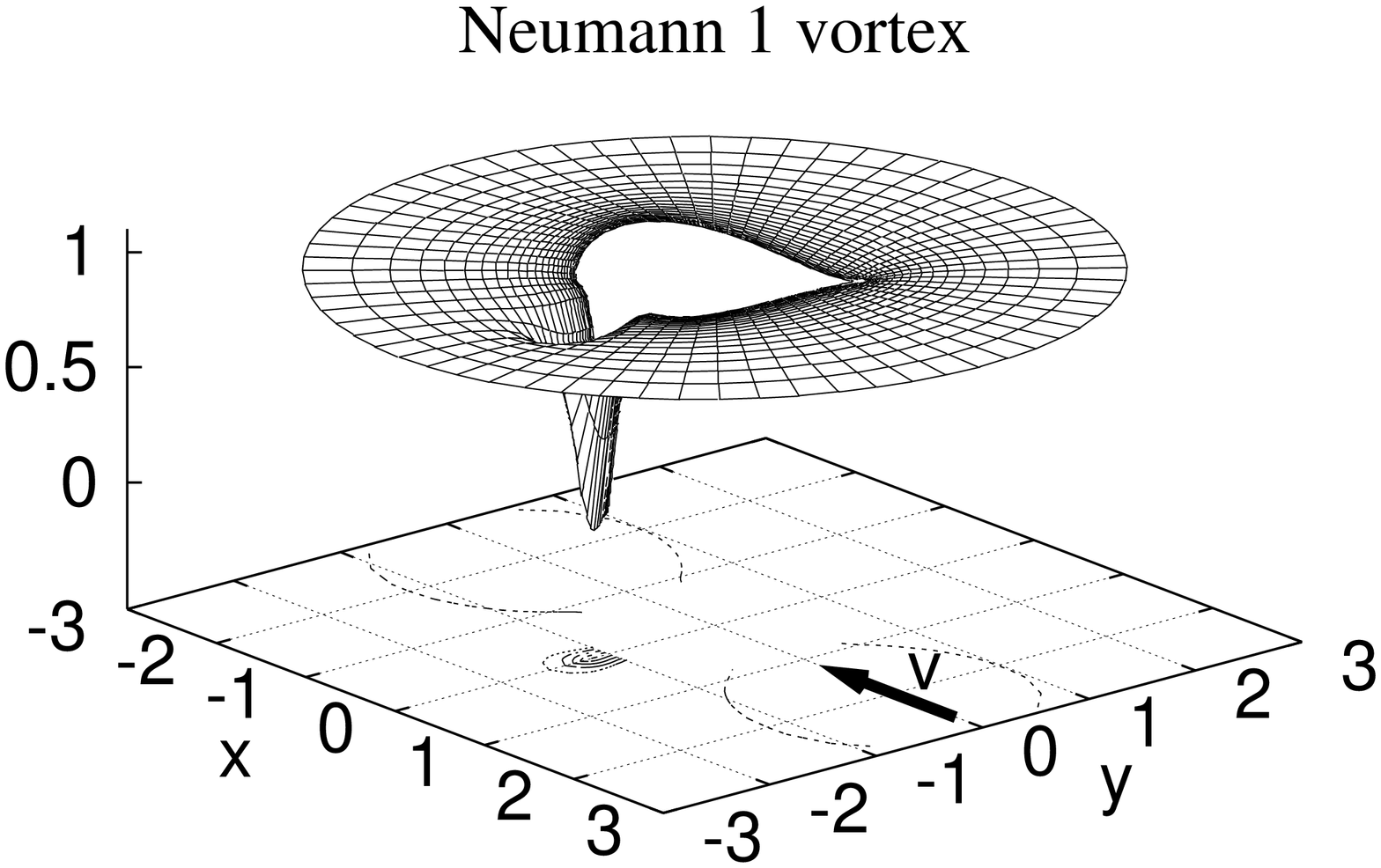}}
\centerline{\includegraphics[width=.4\textwidth, angle=-90]{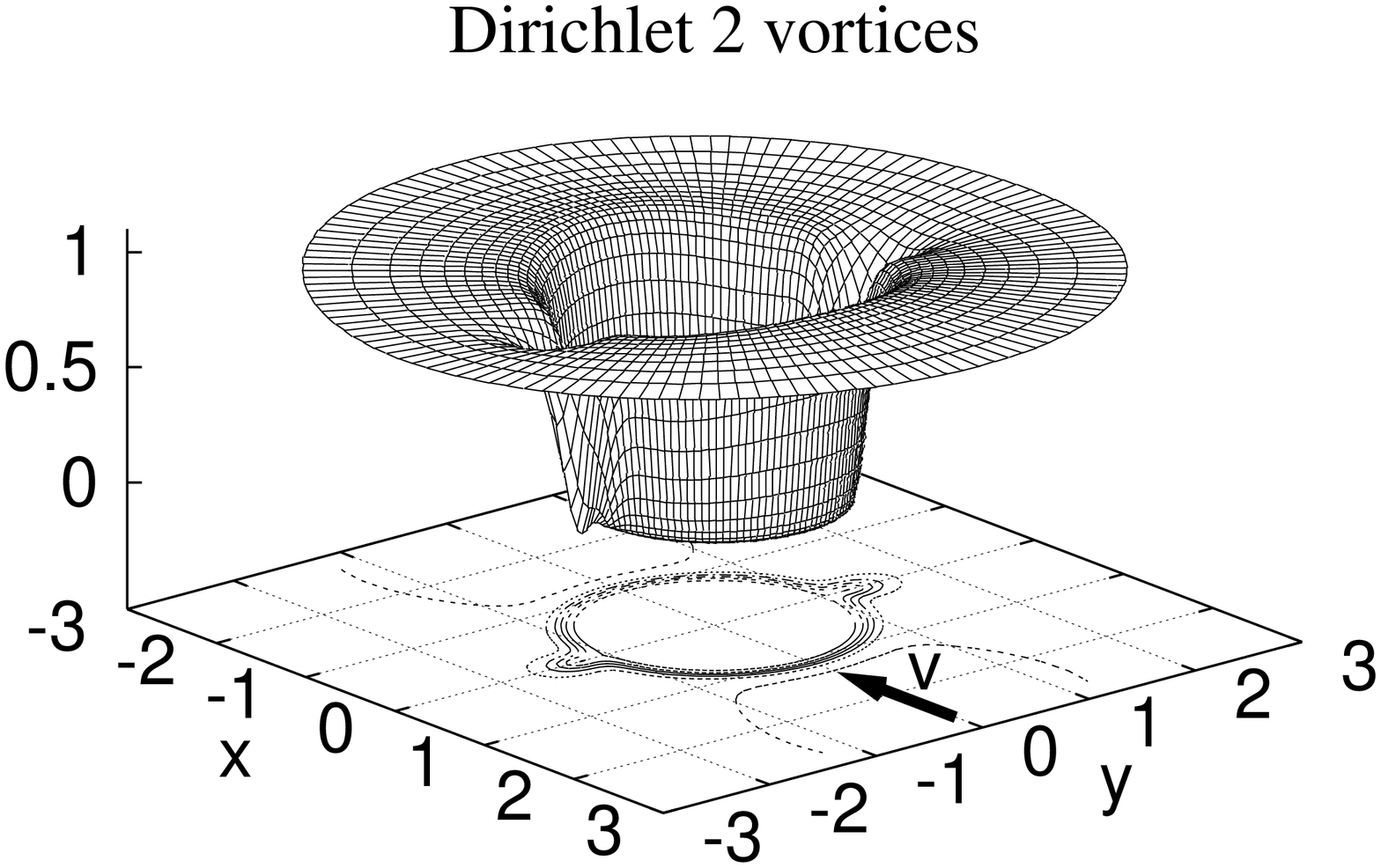}\includegraphics[width=.4\textwidth, angle=-90]{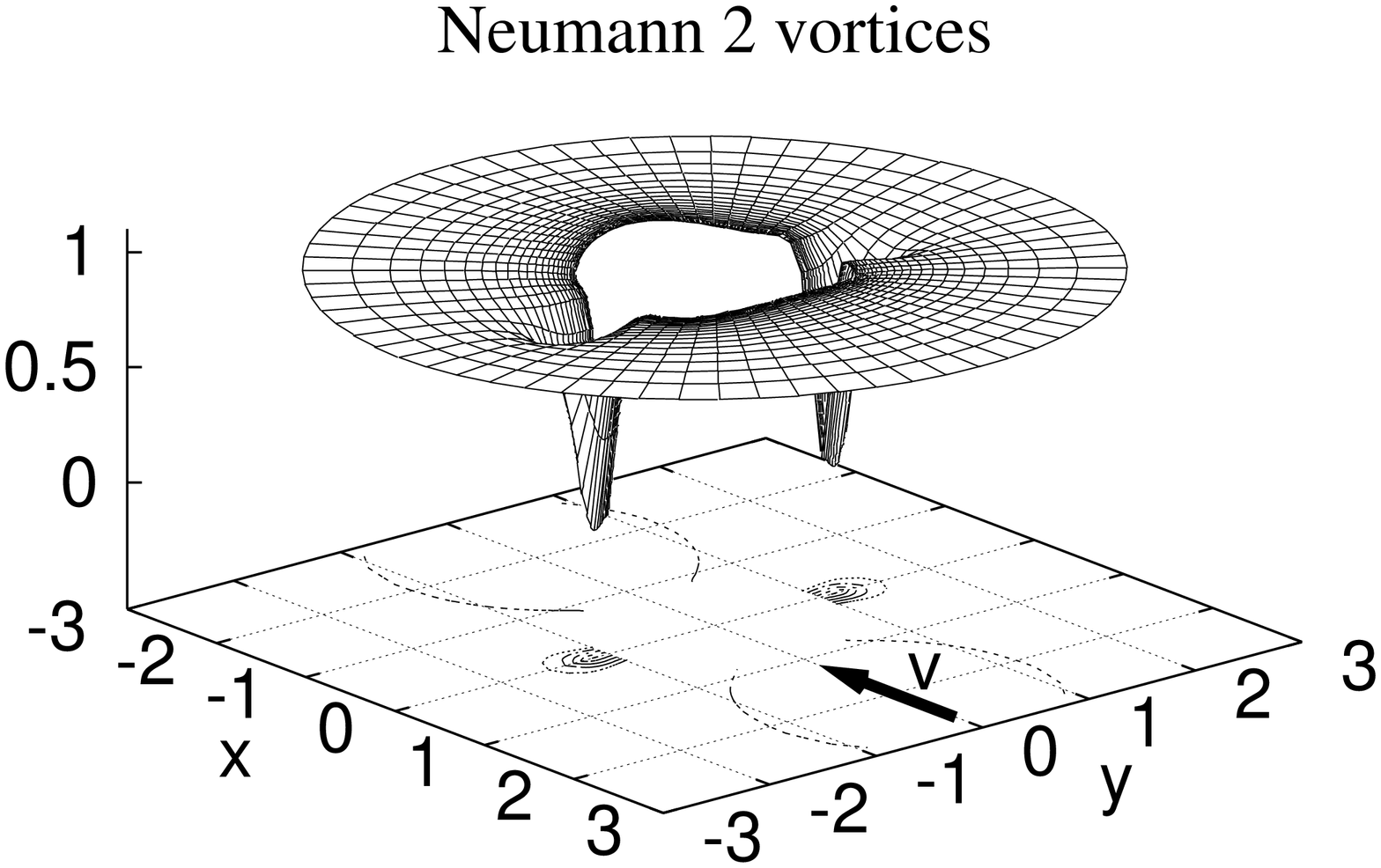}}
\caption{Density $\rho=|\psi|^2$ of stationary solutions for $\xi/D=1/20$ and $\mach=0.3$ far from the bifurcation threshold:
(top) stable solution, (middle) asymmetric unstable solution  and 
(bottom) symmetric unstable solution. Left : Dirichlet conditions; right: Neumann conditions.}
\label{fig:stat_2D_small_xi}
\end{figure}

In Fig.~\ref{fig:nls_2D_diag_bif_small_xi}(a), at a Mach number smaller than $\mach\crit$, the unstable
symmetric branch (dashed line, circle, $\xi/D=1/20$) bifurcates at a pitchfork to a
pair of asymmetric branches (dotted line, $\xi/D=1/20$) \cite{HB97}.
It can be directly checked on our results
(see figure \ref{fig:stat_2D_small_xi}, middle) that the secondary pitchfork bifurcation
breaks the $y \mapsto -y$ symmetry of the flow for both boundary conditions.

At a fixed Mach number, the energy difference between a stable and an unstable
solution corresponds to the energy barrier necessary to dynamically nucleate
an excitation. Note that this barrier for a symmetric unstable solution
is about twice that of an asymmetric unstable solution.


\subsubsection{Stationary solutions}
\label{sec:small_xi}

By visualizing the stationary solutions of the NLSE,
the branches of
Fig.~\ref{fig:nls_2D_diag_bif_small_xi} can be related to the presence of vortices.
Figure~\ref{fig:stat_2D_small_xi} shows the density $\rho=|\psi|^2$ of typical
stationary solutions for $\mach=0.3$ and $\xi/D=1/20$ for the
two types of boundary conditions.
It is apparent by inspection of the figure that
the stable branch is irrotational (figure \ref{fig:stat_2D_small_xi}, top)
while the asymmetric unstable branch corresponds to a one-vortex solution (figure \ref{fig:stat_2D_small_xi}, middle) 
and the symmetric unstable branch,
to a two-vortex solution (figure~\ref{fig:stat_2D_small_xi}, bottom).

For Dirichlet boundary conditions, similar results were found with
periodic pseudo-spectral codes~\cite{HB00}. However, our method
directly imposes the correct boundary conditions without resorting to an artificial repulsive potential.
Also note that the critical Mach number is here determined for a single obstacle,
whereas a periodic array of obstacles was used in previous study. Huepe \etal~\cite{HB00} find
for ratio $\xi/D=1/40$, $\mach\crit^{\mathrm{Huepe}}\simeq 0.3817$ whereas we obtain $\mach\crit \simeq 0.3941$. A single obstacle perturbs less the flow than an infinite array of obstacles (even with large separation), it is therefore natural to find a higher critical Mach number in our simulations.
As $\mach$ is increased, the distance between the vortices and the
obstacle for the unstable branches (figures \ref{fig:stat_2D_small_xi}, middle, bottom,
Dirichlet) decreases.
At a certain $\mach_{nv}<\mach\crit$, the vortices disappear on the surface
on the cylinder, generating an irrotational flow
(see \cite{HB97} for a detailed study of the Mach number at which
one or two vortices emerge from the disk).

Note that the branch following procedure used to compute the unstable branches
bifurcating from the stable branch conserves the velocity circulation. The total velocity circulation around the disk is null.
The two-vortex solution conserves the total circulation since the two vortices
are counter-rotating. For the one-vortex solution, an image vortex located
at the middle of the obstacle has to be invoked. This point will be reconsidered
in section~\ref{sec:large_xi}.

\subsubsection{Variation of critical Mach number with $\xi/D$}

We now study the dependence on $\xi/D$ of the critical Mach number $\mach\crit$.

Our numerical method needs a slight modification to allow us to explore the large $\xi/D$ regime.
The transformation $r(z)$ is modified such that mesh-points situated near
the obstacle are stretched (see section~\ref{sec:new_method}, equation~(\ref{mapgeneralise})). 
This procedure avoids wasting resolution close to the cylinder.

\begin{figure}[ht!]
\centerline{\includegraphics[width=.6\textwidth]{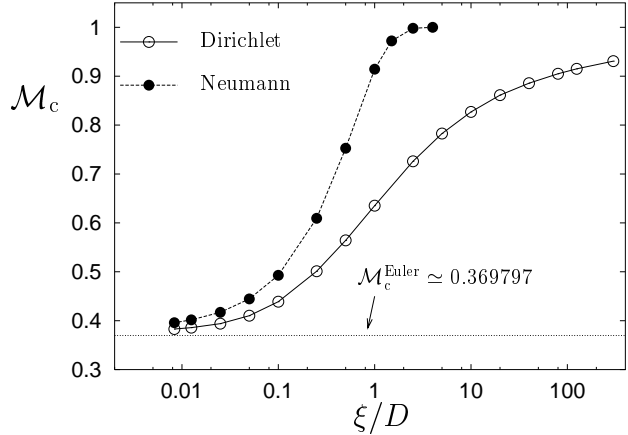}}
\caption{Critical Mach number $\mach\crit$ versus $\xi/D$.
Note that Dirichlet boundary condition solutions admit a smaller $\mach\crit$ than
Neumann boundary condition solutions. They both tend to the Euler critical Mach number
as $\xi/D$ decreases. 
}
\label{fig:nls_2D_mach_critique}
\end{figure}

Results are displayed in figure~\ref{fig:nls_2D_mach_critique}.
For a given type of boundary conditions, the Mach number decreases with $\xi/D$
and, for both boundary conditions, it converges towards the Euler limit for
small $\xi/D$.

First note that the value of the critical Mach number is lower than $1$. As we are interested in stationary solutions with density approaching $1$ at infinity like polynomials in $1/r$ (see sections~\ref{sec:bl} and~\ref{sec:new_method}), the speed of the obstacle $v$ has to remain below the speed of sound $c$. Otherwise radiation of sound waves would occur in the same way as discussed in~\cite{bib:Hakimobstacle,bib:Hakim_Haddad}. 

We now discuss the case of Dirichlet boundary conditions and will extend the
argument to Neumann conditions at the end of this section.

For Dirichlet conditions,
we have seen in section \ref{sec:bl} and in section \ref{sec:num_bl} that, in the case of small $\xi/D$, the boundary layer has a thickness of order $\xi$. The situation at large $\xi/D$ is quite different.

We now show that the effect of the cylinder on the flow at $r>1$ is vanishingly small when $\xi\rightarrow \infty$. At zero Mach number, the density $\rho_\xi(r)=R_\xi^2(r)$ satisfies
\begin{equation}
\xi^2 \left(\partial_{rr}R_\xi+\frac 1 r \partial_r R_\xi \right) - R_\xi^3 + R_\xi = 0 \label{eq:rho_mach0}
\end{equation}
with $R_\xi(1)=0$ and $R_\xi(+\infty)=1$. For large $r/\xi$, $\tilde R_\xi=R_\xi-1$ satisfies after linearization of (\ref{eq:rho_mach0})
\begin{equation}
\xi^2 \left(\partial_{rr}\tilde R_\xi+\frac 1 r \partial_r \tilde R_\xi  \right)- 2 \tilde R_\xi  = 0 \label{eq:rhotikde_mach0}
\end{equation}
Asymptotically for large $r/\xi$, we have $R_\xi(r)\simeq 1 + \tilde R_\xi^{\mathrm{approx}}$ with 
\begin{equation}
\tilde R_\xi^{\mathrm{approx}} = -\mu_\xi \frac{\Kd_0(\frac{\sqrt{2}r}{\xi}) }{\Kd_0(\frac{\sqrt{2}}{\xi})}
\end{equation}
for a given constant $\mu_\xi$. 

Using a shooting method, we have numerically solved Eq. (\ref{eq:rho_mach0}) starting from $r=B\xi$ ($B$ is a sufficiently large constant) with initial conditions $R^{\mathrm{num}}(B\xi) = 1+\tilde R_\xi^{\mathrm{approx}}(B\xi)$ and $\partial_r R^{\mathrm{num}}(B\xi) = \partial_r\tilde R_\xi^{\mathrm{approx}}(B\xi)$ and adjusting the constant $\mu_\xi$ so that $R^{\mathrm{num}}(1)=0$. 

Our shooting method indicates that
\begin{equation}
\lim_{\xi\rightarrow +\infty} \mu_\xi = 1^{+}
\end{equation}

\begin{figure}[ht!]
\centerline{\includegraphics[width=.55\textwidth]{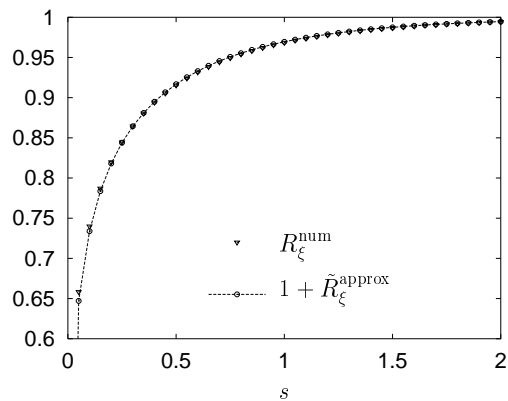}}
\caption{Plot of the function  
$R_\xi\simeq 1 + \tilde R_\xi^{\mathrm{approx}}$ together with $R_\xi^{\mathrm{num}}$ both expressed in the variable $s=(r-1)/\xi$ for $\xi/D = 2500$ (in this case, $\mu_\xi\simeq 1.057$). The function $1 + \tilde R_\xi^{\mathrm{approx}}$ is a very good approximation of the solution of equation (\ref{eq:rho_mach0}) except close to the cylinder ($s=0$).
}\label{fig:plot_rho0}
\end{figure}

Figure \ref{fig:plot_rho0} displays the function $1 + \tilde R_\xi^{\mathrm{approx}}$ together with the numerical solution of Eq. (\ref{eq:rho_mach0}) calculated by the shooting method, expressed in term of the rescaled variable $s=(r-1)/\xi$. Thus, at large $\xi/D$ and for $s>0$, $1 + \tilde R_\xi^{\mathrm{approx}}$ is a very good approximation of the solution of Eq.~(\ref{eq:rho_mach0}). Points close to the cylinder differ, which is obvious since $1 + \tilde R_\xi^{\mathrm{approx}}$ does not vanish at the cylinder. Therefore we have for fixed $s>0$
\begin{equation}
R_\xi(1+\xi s) - 1 \simeq \mu_\xi \frac{\Kd_0(\frac{\sqrt{2}}{\xi}+\sqrt{2}s) }{\Kd_0(\frac{\sqrt{2}}{\xi})} \underset{\xi\rightarrow +\infty}{\sim} -\frac{\Kd_0(\sqrt{2}s)}{\log \xi}
\end{equation}
In this sense, the square-root of the density $R_\xi$ approaches logarithmically the uniform state.
Then the effects of the obstacle on the flow at large $\xi/D$ for Dirichlet conditions  are very small. Thus the critical velocity is expected to increase with $\xi/D$ and the larger $\xi/D$, the closer to $1$ the critical Mach number. 

As the Neumann conditions perturb even less the fluid than the Dirichlet conditions, one can easily understand why the critical Mach number for Neumann conditions increases faster  at large $\xi/D$ than for Dirichlet conditions.

Turning now to the small $\xi/D$ regime, the critical Mach number for Neumann conditions is also found to be larger than that of Dirichlet conditions. This point is quite surprizing since, at small Mach number,  stationary solutions for Neumann conditions approach the Euler stationary states better than the stationary solutions for Dirichlet conditions (see the bifurcation diagrams on figure \ref{fig:nls_2D_diag_bif_small_xi}). However, we can offer the following semi-quantitative argument. The critical Mach number decreases with decreasing $\xi/D$. We have shown in section \ref{sec:bl} that the effective radius $r\eff(\xi)$ of stationary NLS flows at small Mach numbers was bigger for Dirichlet conditions than for Neumann conditions. Assuming that this result holds for bigger Mach numbers (of order $\mach\crit^{\mathrm{Euler}}$), one can consider that the Neumann conditions stationary solutions have the same critical Mach number as the Dirichlet stationary solutions when they reach the same ratio $\xi/D\eff(\xi)$, imposing therefore smaller values of $\xi/D$ in the Neumann case.

Finally, note that we have found no numerical indication showing that \linebreak $\mach\crit(\text{Dirichlet})$ could become bigger that $\mach\crit(\text{Neumann})$, for very small values of $\xi/D$.

\subsection{Bifurcation diagrams and stationary states at large coherence length}
\label{sec:large_xi}

The large $\xi/D$ regime could be reached experimentally by considering BEC with large coherence lengths perturbed by a sharply focused detuned laser. 
As seen in the previous section, the critical Mach number tends very quickly towards $1$ for Neumann boundary conditions.
Furthermore, these conditions are academic and have no experimental equivalent in BEC. 
Thus, we will study the limit of large $\xi/D$ only for Dirichlet conditions, the experimentally realistic ones. 

\begin{figure}[ht!]
\centerline{ \includegraphics[height=0.2\textheight]{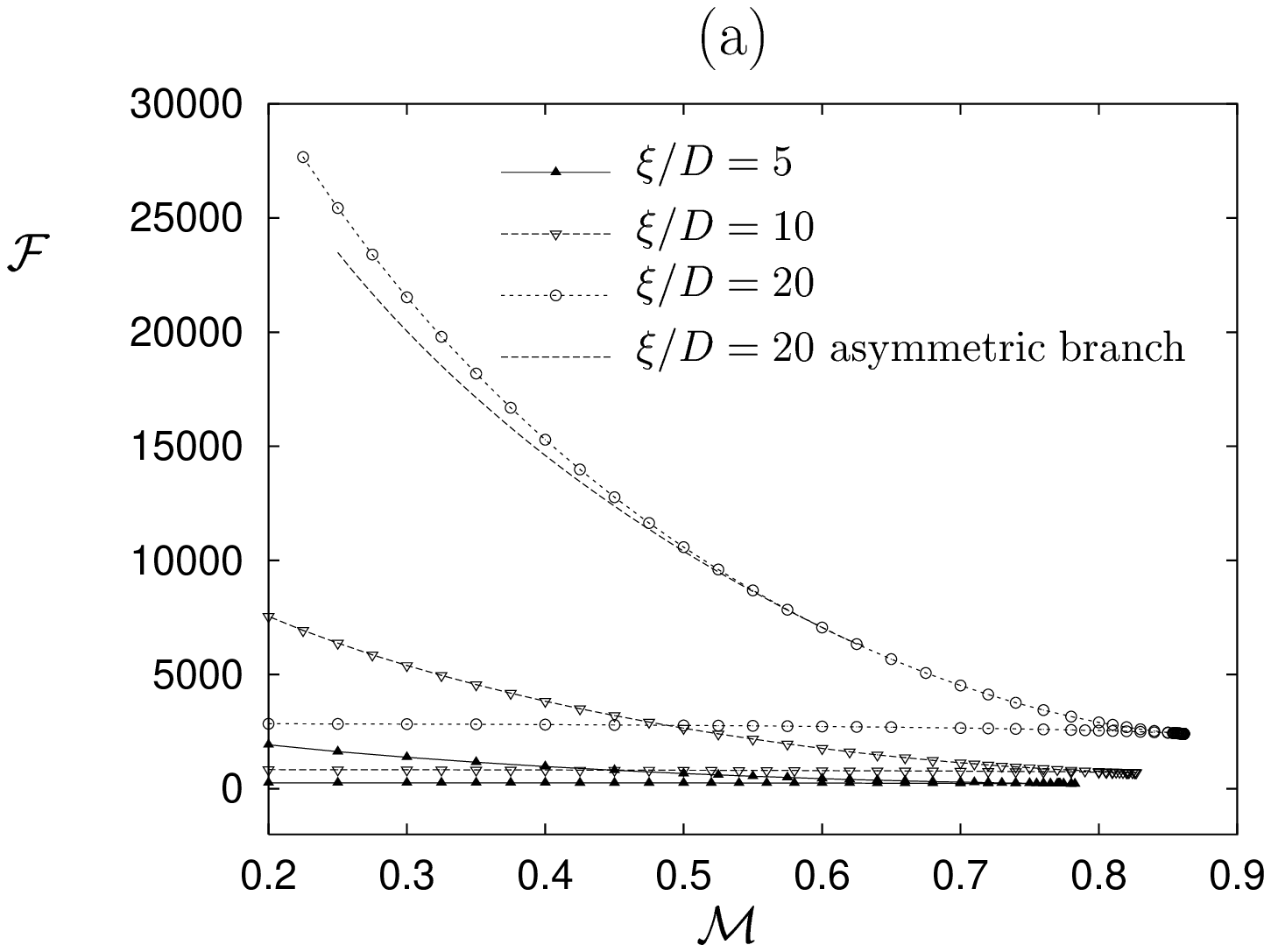}\hfill  \includegraphics[height=0.2\textheight]{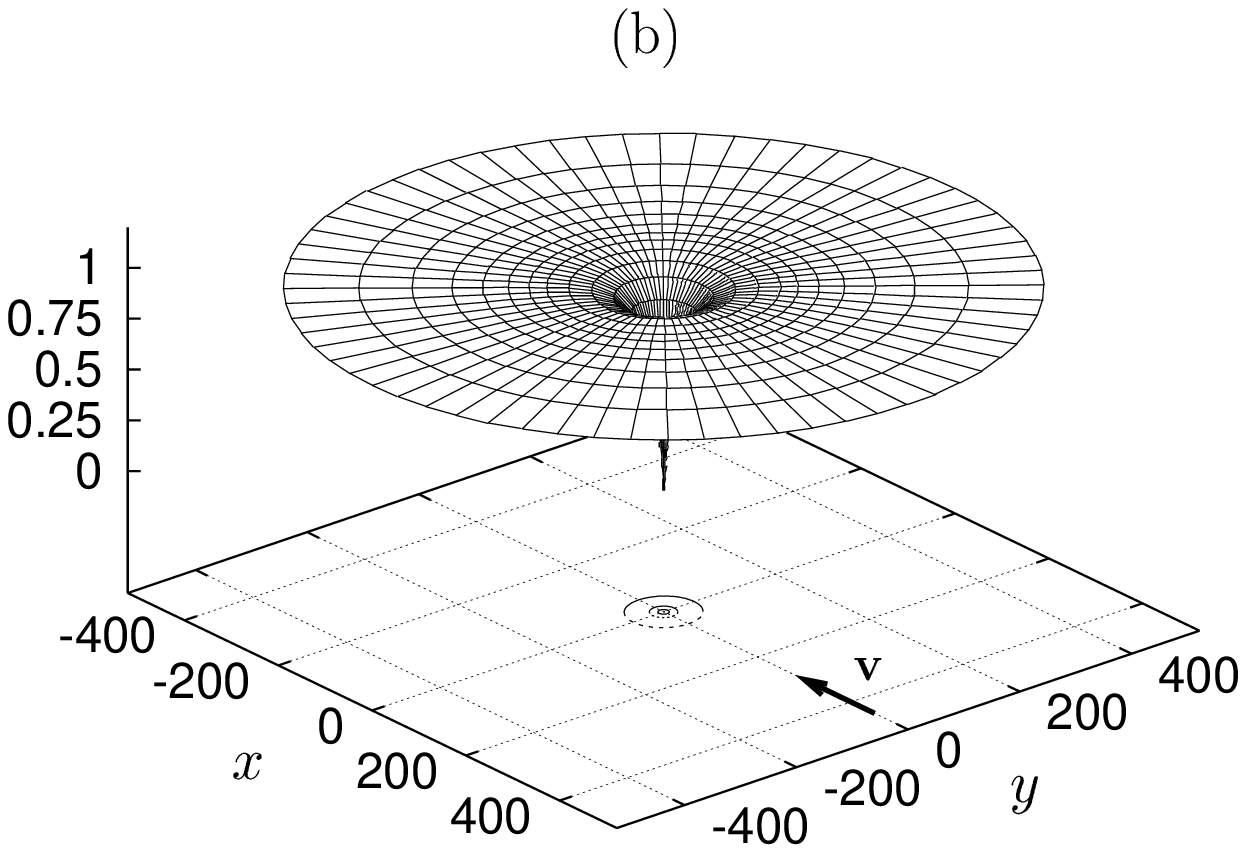}}
\centerline{ \includegraphics[height=0.2\textheight]{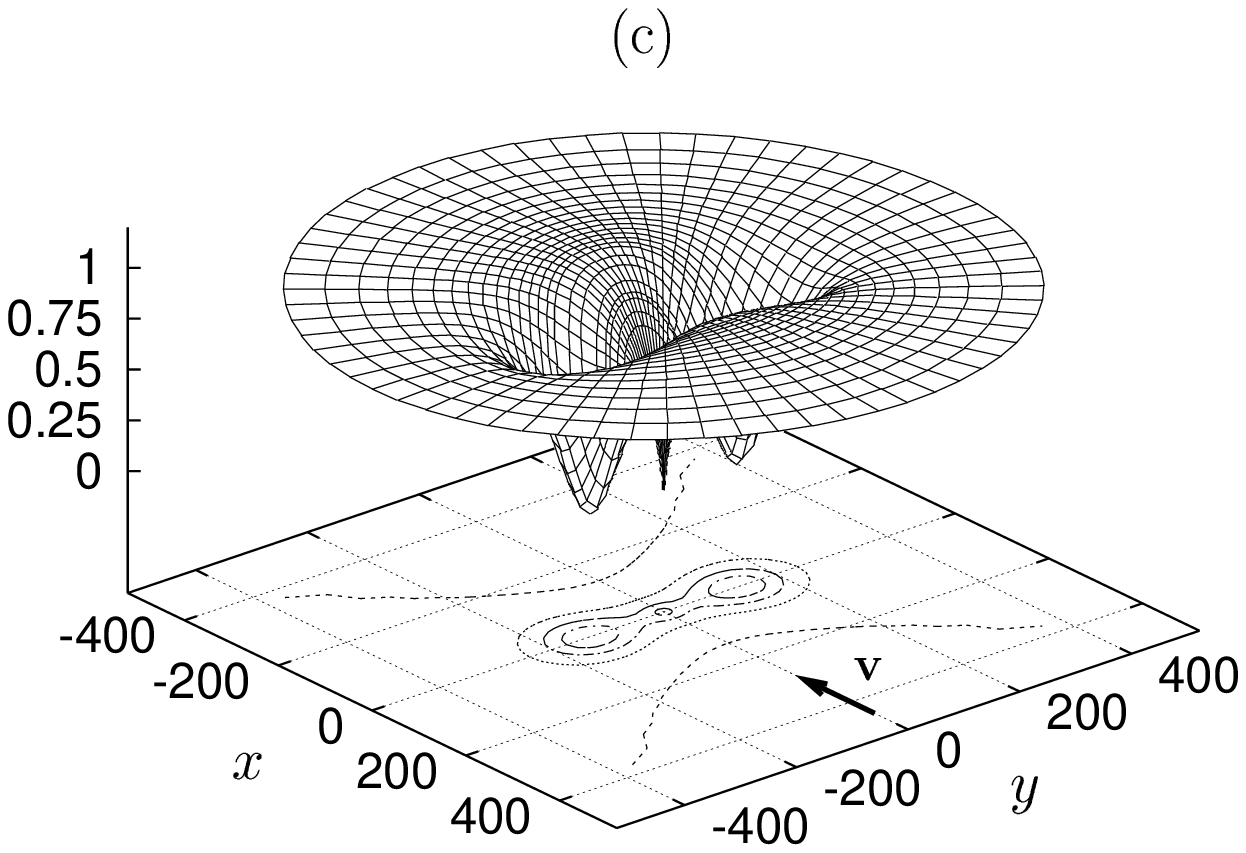}\hfill
 \includegraphics[height=0.2\textheight]{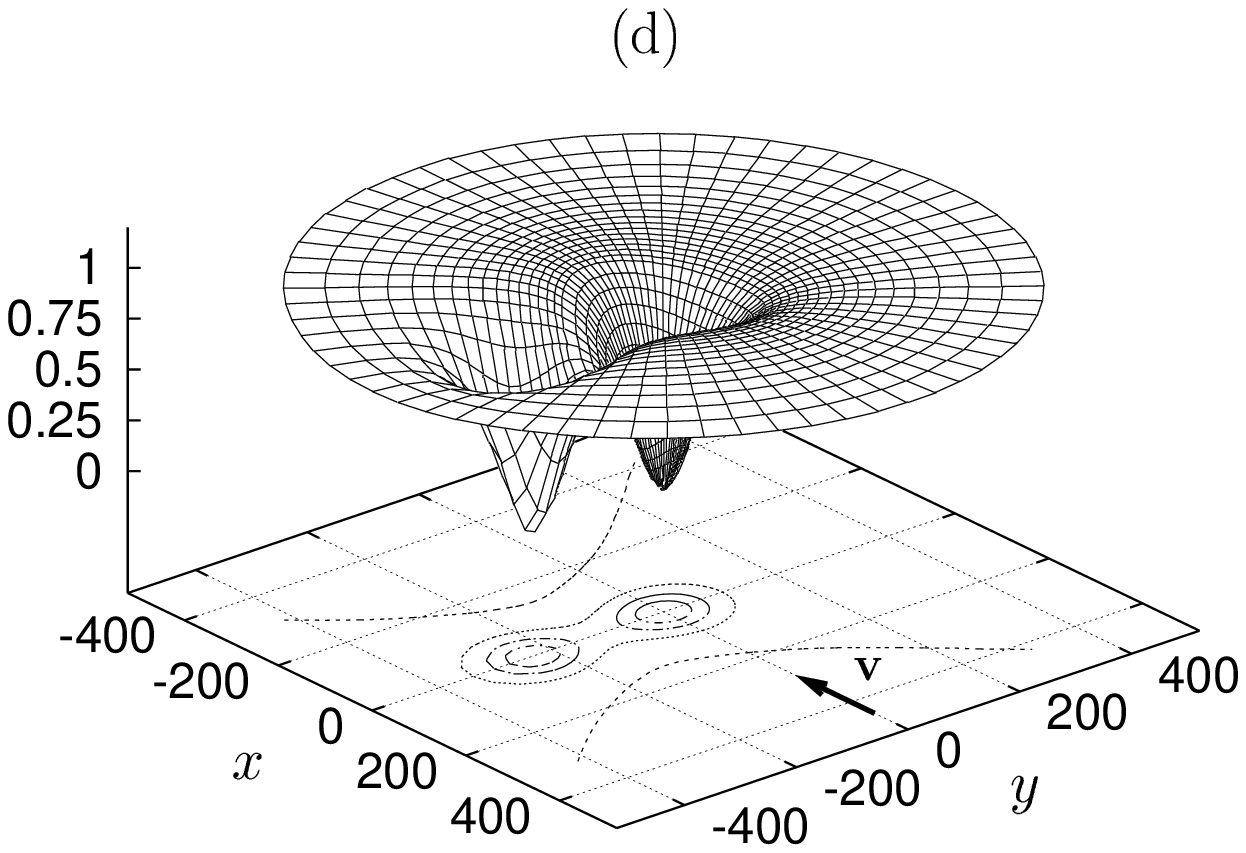}}
\caption{(a) Bifurcation diagram for large $\xi/D$ for Dirichlet conditions. 
Density of stationary solutions for $\xi/D=20$ and $\mach=0.25$ far from the bifurcation threshold  ($\mach\crit \simeq 0.86$):
(b) stable solution, (c) symmetric unstable solution  and 
(d) asymmetric unstable solution. }
\label{fig:nls_2D_diag_bif_large_xi}
\end{figure}

The bifurcation diagram, computed for Dirichlet boundary conditions
and different $\xi/D$, is displayed on figure~\ref{fig:nls_2D_diag_bif_large_xi}(a).
Just like in the small $\xi/D$ case, a branch of stable solutions is connected to
a branch of unstable solutions through a saddle-node bifurcation
at a critical Mach number $\mach\crit$. The values of $\mach\crit$ are seen  to approach $1$
when $\xi/D$ increases.
The corresponding stationary solutions are also displayed on figure~\ref{fig:nls_2D_diag_bif_large_xi}.
The stable solution (b) is irrotational while the two unstable solutions contain respectively
one (d) or two vortices (c) far from the critical Mach number.
However, close enough to the bifurcation tip, the unstable stationary solution shows no 
$2\pi$ phase jump and therefore no vortices are present. 
We will come back to this point in the next section.

As already pointed in section~\ref{sec:small_xi}, the one-vortex solution 
that breaks the symmetry $y\mapsto -y$ associates
a vortex located outside the obstacle to an image vortex situated inside the obstacle.
The image vortex is clearly visible on figure~\ref{fig:nls_2D_diag_bif_large_xi}(d)
because its core is larger than the obstacle itself (compare with figure~\ref{fig:nls_2D_diag_bif_large_xi}b or c).
Note that the large $\xi/D$ energy difference between the stable and the
asymmetric solution branches is not half the energy difference between the stable and the
symmetric solution branches, contrary to the small $\xi/D$ case.  This stems from the fact that, for large $\xi/D$, the energy of the asymmetric branch also includes the additional contribution of the now visible image vortex.

%
%
%
%
%

\section{Dynamical results}
\label{sec:dyn}

Solutions of the NLSE (\emph{in the absence of an obstacle}) in dimension $2$, moving at constant speed while preserving their shape, have been exhibited by Roberts \etal~\cite{bib:Jones4,bib:Jones5}. These solutions are pairs of counterrotating vortices but also what they called rarefaction pulse (depletion pulse with non zero density and therefore no vorticity).

A natural question is then to know which kind of excitations can be nucleated past a disk.

\subsection{Nucleation of vortices}

The stationary solutions obtained numerically provide us 
with adequate initial data for the study of dynamical solutions.
Indeed, after a small perturbation, their integration in time
will generate a dynamical evolution with
very small acoustic emission. This procedure also provides an
efficient way to start vortical dynamics in a controlled manner.

\begin{figure}[ht!]
\centerline{(a)\includegraphics[width=.35\textwidth, angle=-90]{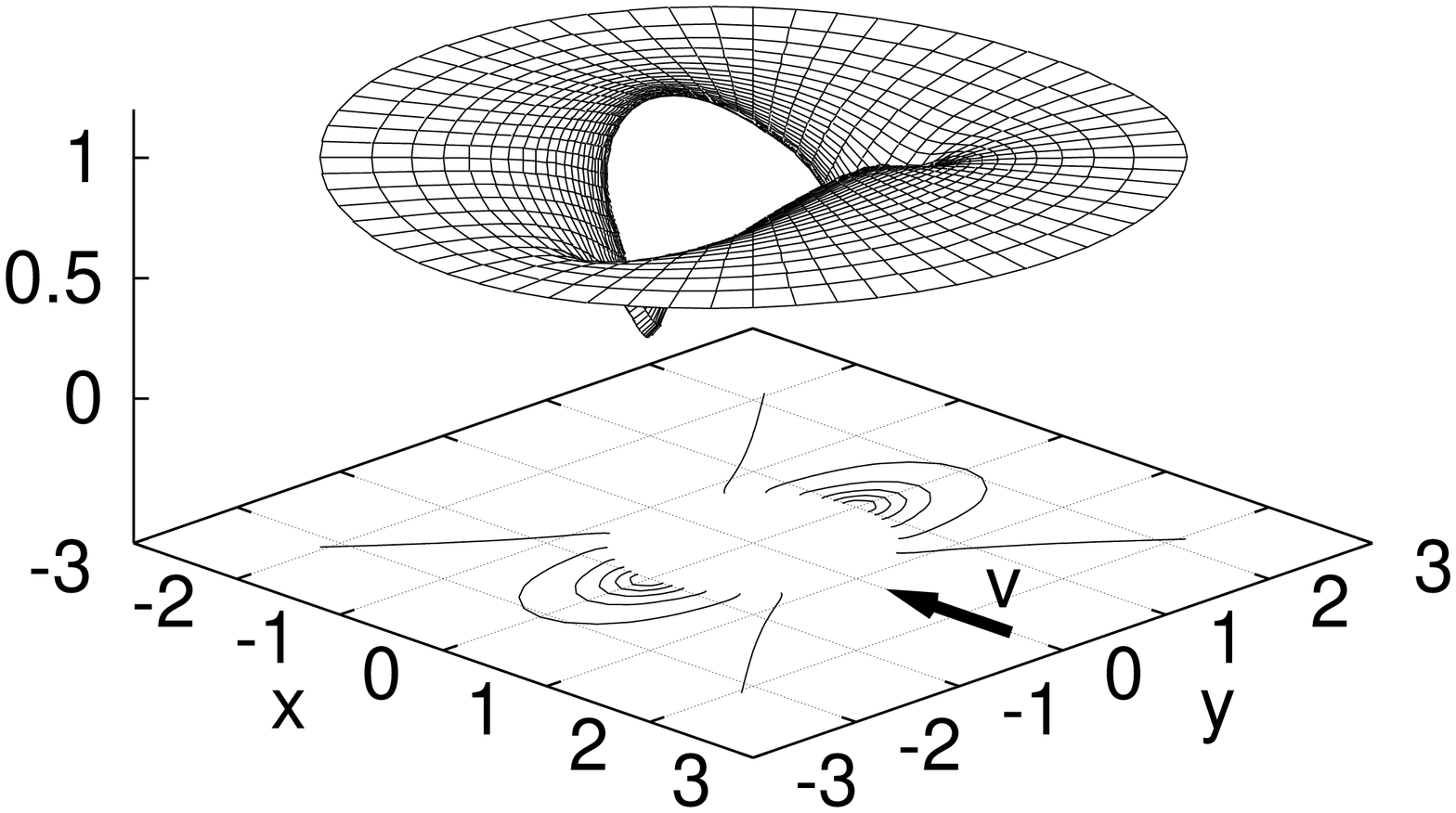}
\qquad (b) \includegraphics[width=.35\textwidth, angle=-90]{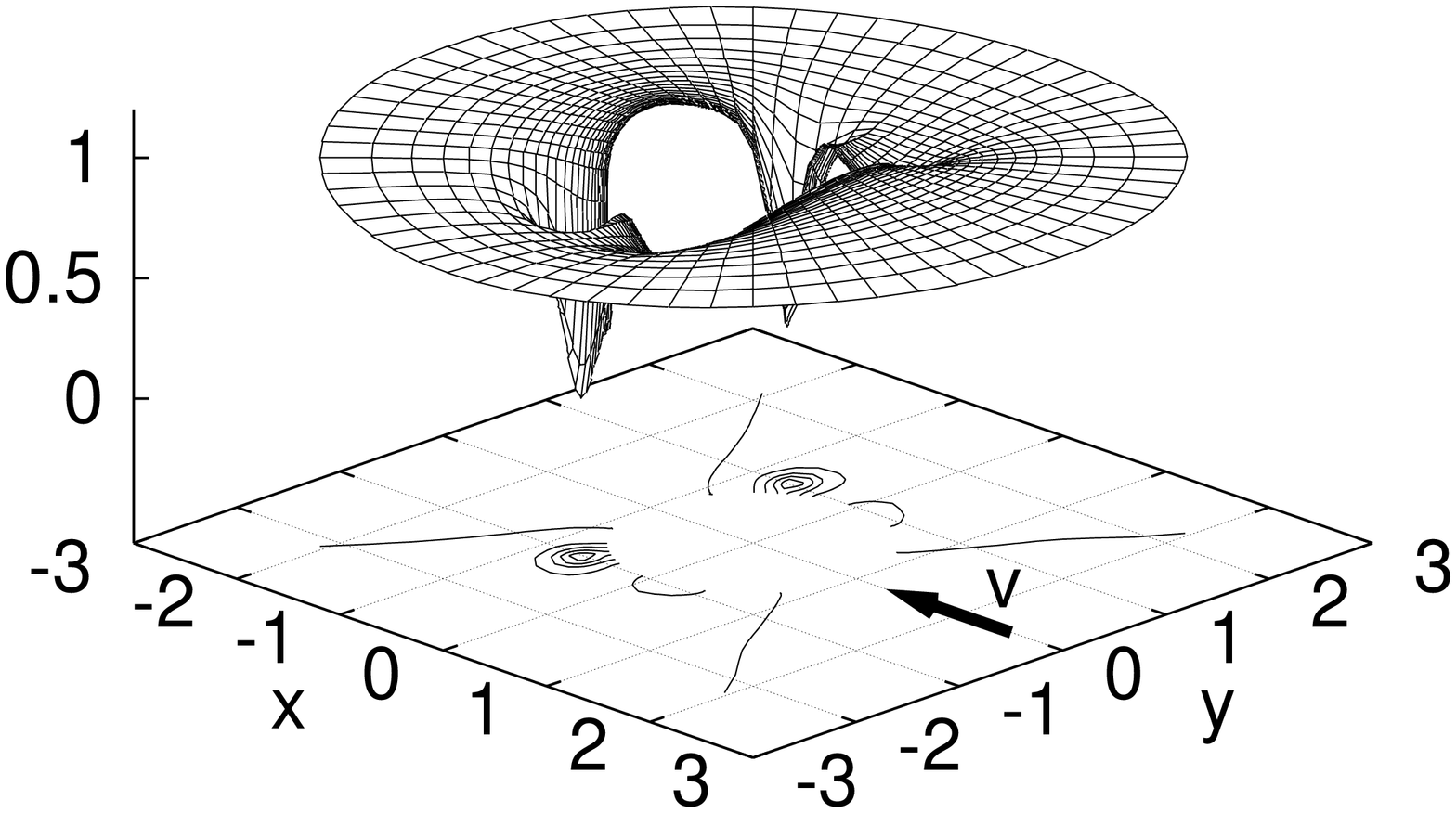}}
\centerline{(c)\includegraphics[width=.35\textwidth, angle=-90]{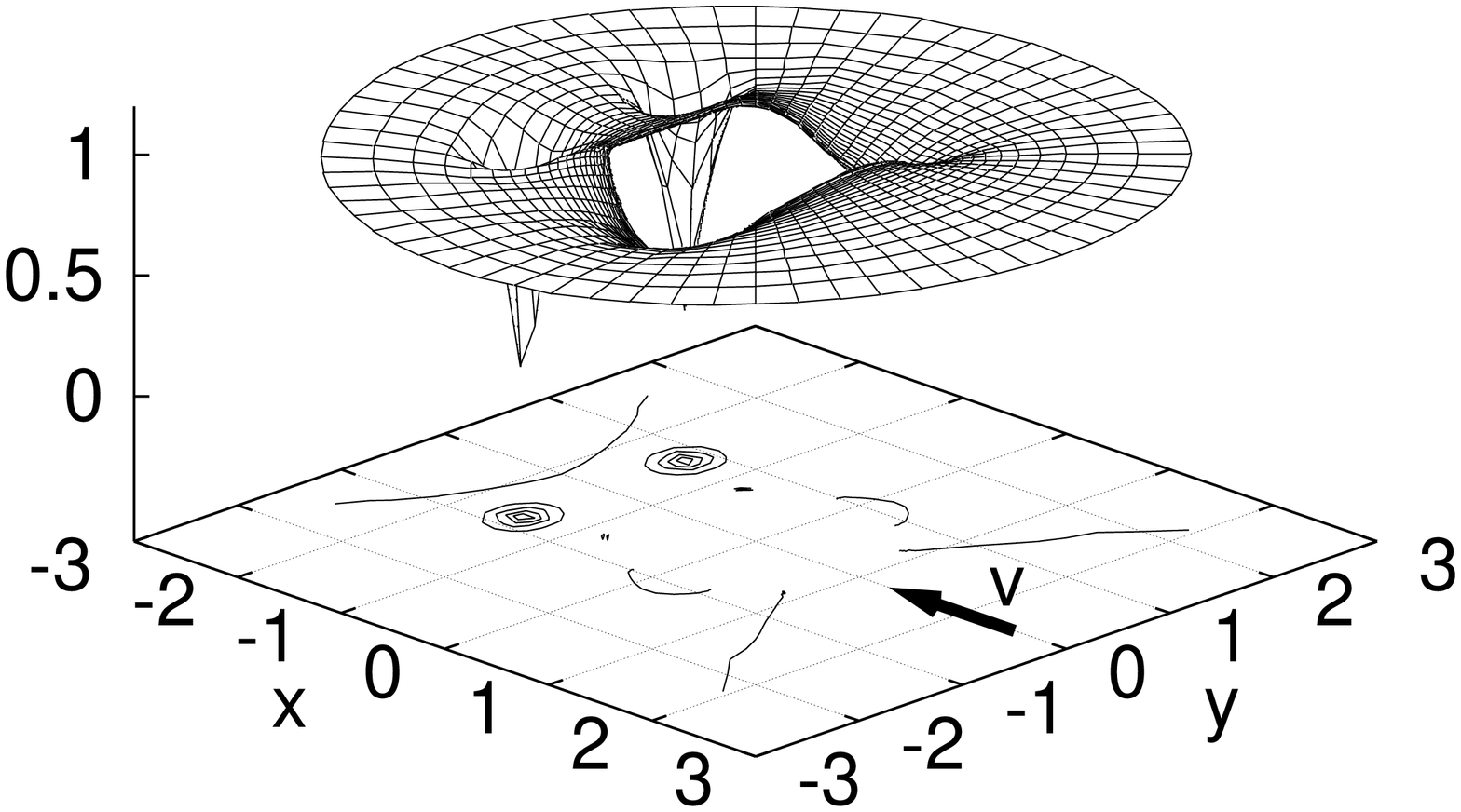}
\qquad (d) \includegraphics[width=.35\textwidth, angle=-90]{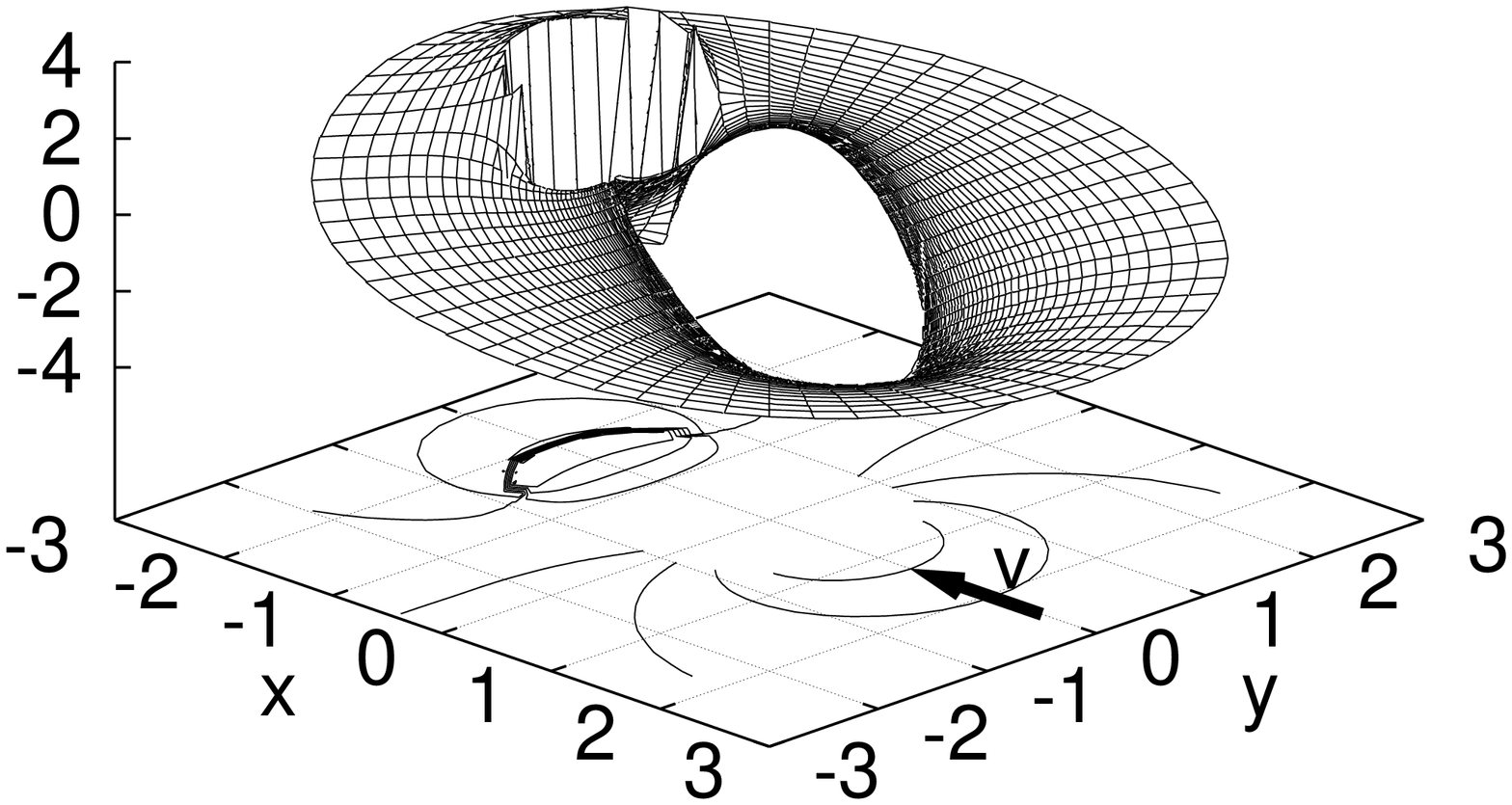}}
\caption{Typical vortex nucleation in the case of Neumann boundary conditions, for $\xi/D=1/20$. (a,b,c) Plot of the density at different times (in arbitrary unit $t=0$, $t=230$, $t=260$ respectively). (d) At $t=260$, phase of the system. Note the $2\pi$-phase jumps around the two points where the density vanishes:  iso-phase lines emerge from these two points.}
\label{fig:nls_2D_nucleation de vortex}
\end{figure}

It is already known from studies performed using a repulsive potential (see figure 6
of \cite{HB00}) that, at small values of $\xi/D$, vortex pairs are dynamically nucleated.
The same behavior is obtained using the present numerical method with Dirichlet boundary conditions 
(data not shown).
This behavior persists when using Neumann boundary conditions, as shown on figure~\ref{fig:nls_2D_nucleation de vortex} 
that displays the nucleation of a
$+$ and $-$ vortex pair (which will be followed by a periodic
emission of other pairs). 
The phase of the complex field exhibits a $2\pi$ jump for each vortex
(see figure~\ref{fig:nls_2D_nucleation de vortex}(d)).

\subsection{Nucleation of rarefaction pulse}

For large $\xi/D$, using Dirichlet boundary conditions, we have proceeded in the same way as in the previous subsection by perturbing an unstable symmetric solution at Mach number $\mach > \mach\crit$ to observe the nature of the nucleated excitations.  
The behavior is somewhat more complicated. We show on Table~\ref{tab:diag_phase} the nature of the emitted excitations as a function of $\xi/D$ and $\mach/\mach\crit$. For $\xi/D>15$, and obstacle speed above $\mach\crit$,
a rarefaction pulse (RP) is dynamically obtained; for $\xi/D < 15$, there exists a threshold in $\mach$ under which vortex pair rather than a rarefaction pulse emerges. In some cases (BL for border line), it is unclear whether we have an emitted rarefaction pulse or a vortex pair (the density minimum in such limit cases approaches zero and there is a strong variation of the phase). By measuring the borderline  speed of translation  $\upsilon$ (in the frame at rest) of the emitted excitation (see Table~\ref{tab:vitesse}), we found that it is very close to the known limit speed of translation $ \upsilon_{\mathrm{V}}$ at which occurs the change in the nature of excitations in 2D superflow in the absence of an obstacle: $\upsilon_{\mathrm{V}}/c = 0.43 \sqrt 2 \simeq 0.61$~\cite{bib:Jones5}.

\begin{figure}[ht!]
\centerline{{ (a)}
\includegraphics[width=.33\textwidth, angle=-90]{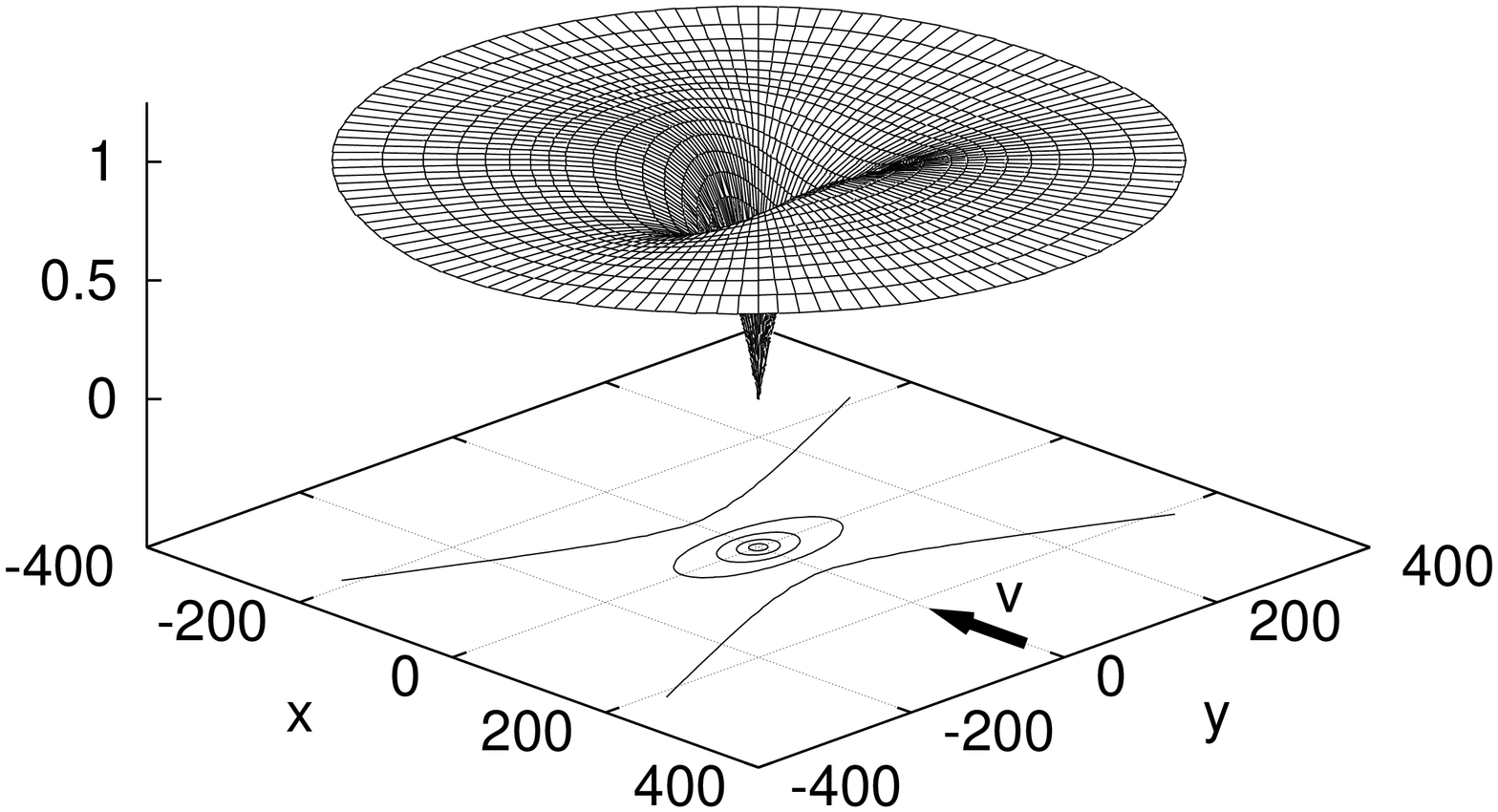}
{ (b)}
\includegraphics[width=.33\textwidth, angle=-90]{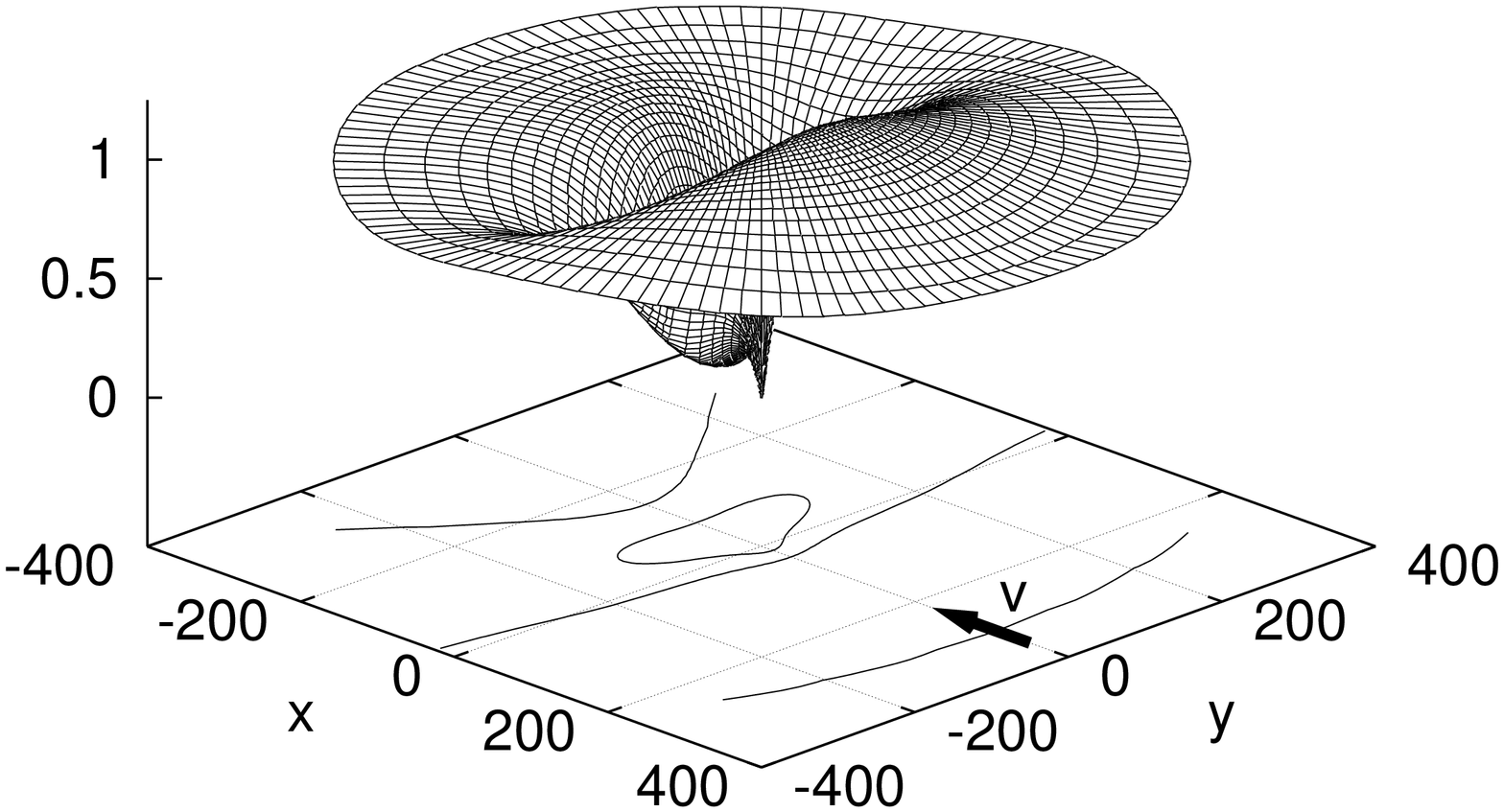}}

\centerline{{ (c)}
\includegraphics[width=.33\textwidth, angle=-90]{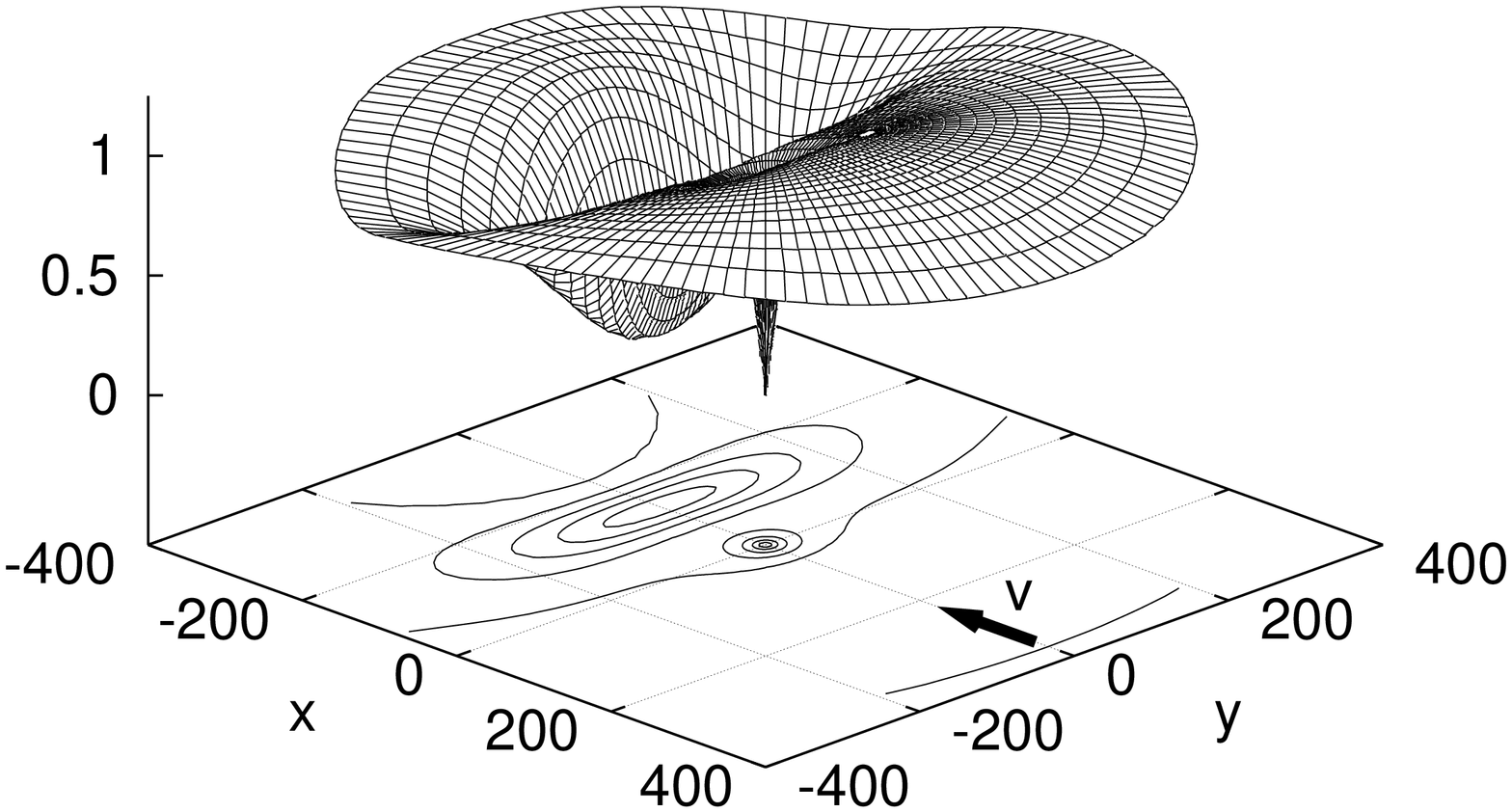}
{ (d)}
\includegraphics[width=.33\textwidth, angle=-90]{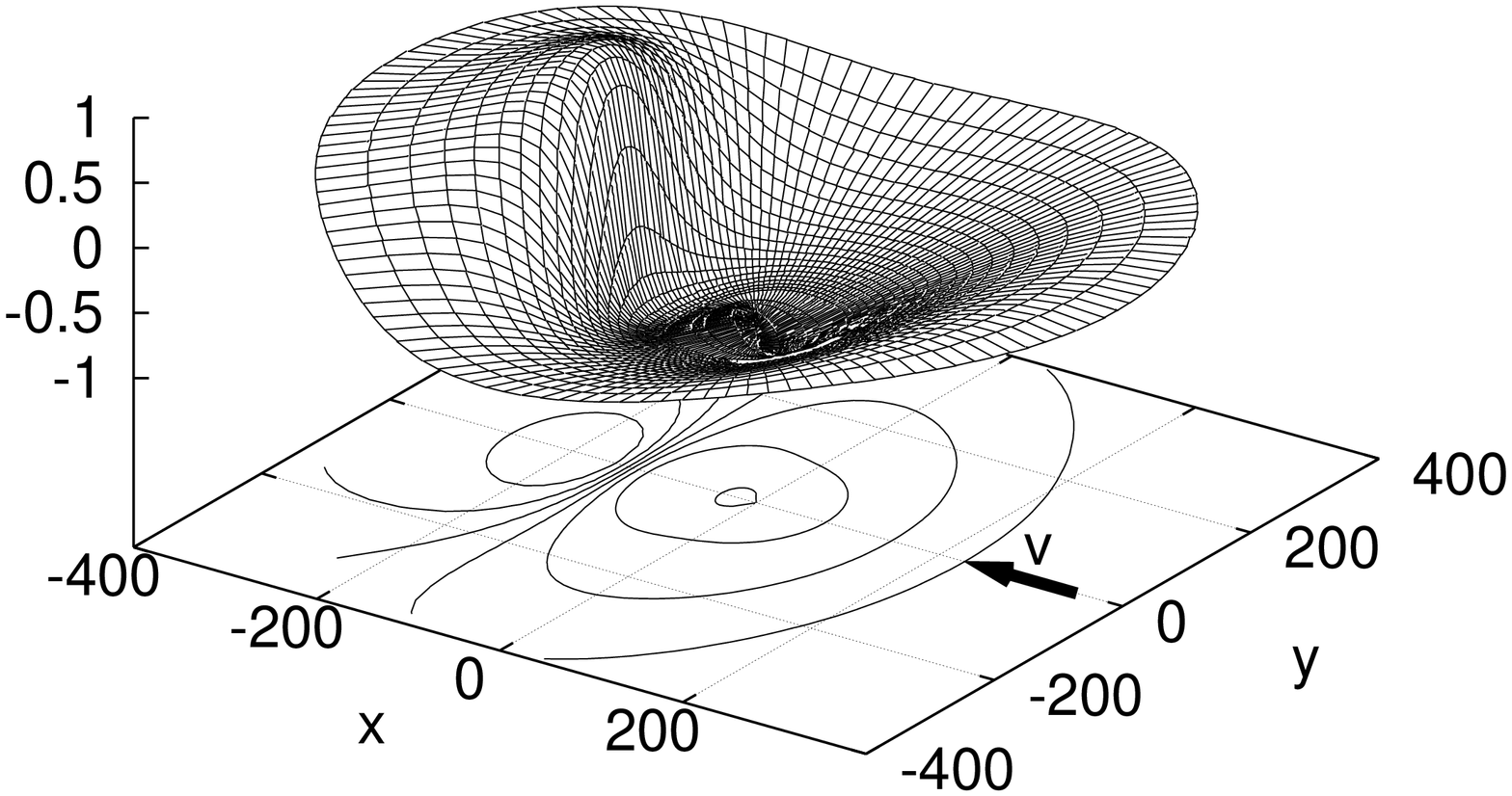}
}
\caption{Nucleation of a rarefaction pulse with Dirichlet boundary conditions for $\xi/D=17.5$ and $\mach/\mach\crit = 1.1$:
density at different times (in arbitrary unit) (a) $t=0$, (b) $t=150$, (c) $t=220$.
(d) Plot of the phase at $t=220$. There is no phase jump, hence no vorticity. The speed of translation of this rarefaction pulse in frame at rest is $0.80$ and its minimum of density is approximately $0.081$.}
\label{fig:nls_2D_nucleation_soliton_gris}
\end{figure}

A desexcitation of an unstable stationary solution at $\xi/D=17.5$ creates
a rarefaction pulse as shown in figure~\ref{fig:nls_2D_nucleation_soliton_gris}. The minimum of density of such a pulse is non zero (here the minimum equals approximately $0.081$), and no phase jump is present (see figure~\ref{fig:nls_2D_nucleation_soliton_gris}(d)).

The change in the nature of the excitation can be understood by the following qualitative argument. A rarefaction pulse can be seen as the superposition of a pair of vortices so close to each other that the minimum density is non zero. 
Close to the critical Mach number, no vortex is detached from the disk. When nucleated, these vortices follow the boundary of the obstacle and then leave it separated by a distance of order $D$. For large $\xi$, the vortices are so close that they give rise to a rarefaction pulse.

We did not study the periodic emission of rarefaction pulses at supercritical regime as done in previous studies~\cite{HB00}. Our numerical method is not adapted for such problems: the mesh is more and more stretched far from the obstacle so that one lacks resolution at long distance to resolve the nucleated excitations.

\begin{table}[ht!]
\begin{center}
\begin{tabular}{|c|c|c|c|c|c|c|}
\cline{3-7}
\multicolumn{2}{c|}{}&\multicolumn{5}{c|}{$  \xi / D$}\\
\cline{3-7}
\multicolumn{2}{c|}{}&$  7.5$&$10$&$12.5$&$15$&$17.5$\\
\hline
      & $1.2$  &BL&RP&RP&RP&RP \\
\cline{2-7}
       &$1.15$&  VP  &RP & RP  & RP & RP \\
\cline{2-7}
 $\frac{\mach}{\mach\crit}$ & $1.1$ &  VP   & BL & RP & RP & RP \\
\cline{2-7}
  & $1.05$ & VP    &VP & VP  &RP  & RP  \\
\cline{2-7}
  & $1.01$ &  VP   & VP& VP & BL  & RP   \\
\hline 
\end{tabular}
\end{center}
\caption{Phase diagram of the nature of emitted excitations as a function of the ratio $\xi/D$ and the Mach number normalized by the critical Mach number. VP and RP respectively stand for vortex pair and rarefaction pulse. BL stands for limit cases where it is hard to distinguish the exact nature of the excitation (the density minimum is very close to zero and the phase has a strong variation).}\label{tab:diag_phase}
\end{table}

\begin{table}[ht!]
\begin{center}
\begin{tabular}{c|ccccc}
$\xi/D$&$  7.5$&$10$&$12.5$&$15$&$17.5$\\
\hline
$\upsilon/c$ &  0.56   & 0.59 & 0.77 & 0.70 & 0.80
\end{tabular}
\end{center}
\caption{Speed of translation $ \upsilon/c$ of the nucleated excitations at $\mach/\mach\crit=1.1$ as a function of $\xi/D$. For the limit case $\xi/D=10$, we find $\upsilon\simeq 0.59$; the change in the nature of excitation in a 2D superflow without obstacle appears at speed equal to $\upsilon_{\mathrm{V}}/c\simeq0.61$~\cite{bib:Jones5}.}\label{tab:vitesse}
\end{table}

%
\section{Conclusion}
\label{sec:conclusion}

The main virtue of the pseudo-spectral method  that we have used in the present study
is its ability to accurately accommodate both large-$r$ asymptotic behavior and thin boundary layers near the cylindrical obstacle, at $r=1$.
Indeed, using modest resolutions, we were able to obtain the Eulerian critical Mach number with $11$
significant digits. In the NLSE case, spectral convergence was obtained on the whole bifurcation diagram for values of $\xi/D$ as low as  $1/120$.

Small coherence length boundary-layer approximations to the stationary solutions were calculated. 
These analytical results were found to be in very good agreement with the numerical results.
The long-range contribution was physically interpreted as a renormalization of the diameter of the disk.

As a by product of our new method, we were able to investigate
not only the physically realistic case of Dirichlet boundary conditions
but also the more academic case of Neuman conditions. 
The influence of the boundary conditions on the stationary solutions of the problem, 
especially their effects on the boundary layer and on the critical Mach number, were investigated. 

For Dirichlet boundary conditions the qualitative results previously obtained, using
periodic pseudo-spectral codes~\cite{HB00}, were recovered. However, our new method
directly imposes the correct boundary conditions, without resorting to an artificial repulsive potential.
Also, the newly obtained critical Mach number is here determined for a single obstacle,
whereas a periodic array of obstacles was considered in previous studies.
Thus, the present article presents the first precise quantitative determination of the critical Mach number
as a function of $\xi/D$ in this reference problem. 

Finally we were able to show  that a transition occurs in the nature of the emitted excitation at large coherence length.
For $\xi/D>15$, the nucleated excitations are rarefaction pulses  whereas, at $\xi/D < 15$, both vortices and rarefaction pulses can be obtained.

\section*{Acknowledgments}

The computations were carried out on the NEC-SX5
computer of the Institut du D\'eveloppement et des Ressources en
Informatique Scientifique (IDRIS) of the Centre National pour la Recherche
Scientifique (CNRS), project no. 0283.


%
%
%
%
\appendix
%
%
\section{Numerical Methods}\label{sec:app_num_meth}
We detail here the numerical procedures used in the simulations
together with the numerical convergences with spatial resolutions
for the critical Mach number and the stationary solutions.
%
%
%
\subsection{Spectra}

In this article, we define the $r$-spectrum and $\theta$-spectrum of a field $\psi$ spectrally represented by $\psi_{n,p}$ as the respective sequence of numbers
\begin{align}
&\mathrm{Sp}^r(p) =\sum_ {n=-\frac{N_\theta} 2 +1}^{\frac{N_\theta} 2} |\psi_{n,p}|^2  && 0\leq p \leq N_r\\
&\mathrm{Sp}^\theta (n)=\sum_ {p=0}^{N_r}|\psi_{n,p}|^2 &&  0\leq n\leq \frac{N_\theta}{2}
\end{align}
Only half the $\theta$-spectrum is considered for the $\theta$-representation is complex-conjugated.

\subsection{Implementation of the boundary conditions}

Contrary to the analytical computations, we work with the complex variable  $\psi$ in order to take account of the possible presence of vortices. In order to impose the boundary conditions, we set
\begin{align}
&\Phi_0(\theta,r) = v\frac{r_0^2\cos\theta}{\sqrt 2 c\xi r}\\
&\psi\new = \psi\old \ed^{\id\Phi_0}
\end{align}
Note that this change of variables does not affect the Dirichlet conditions that still read
\begin{align}
\psi\new|_{\partial\Omega} = 0 \label{bc_NLSDirichlet}
\end{align}
With these new variables, Neumann conditions read
\begin{align}
\partial_r\psi\new|_{\partial\Omega} = 0\label{bc_NLSNeumann}
\end{align}
and the NLSE turns into
\begin{multline}
\id \partial_t\psi\new = \frac{c}{\sqrt{2} \xi}\left[-\xi^2\Delta\psi\new + (|\psi\new|^2-1)\psi\new  \right] + \id\vvec\cdot\nabla\psi\new \\
+ \frac{c}{\sqrt{2} \xi}\left[ \xi^2 (\nabla\Phi_0)^2\psi\new + \xi^2 2\id (\nabla\Phi_0)\nabla\psi\new\right] - \vvec\cdot(\nabla\Phi_0)\psi\new
\end{multline}
with $\psi\new|_{r=1}=0$ (Dirichlet) or $\partial_r\psi\new|_{r=1}=0$ (Neumann). 

\subsection{Time steppings}
\subsubsection{Stationary States}

We search for stationary solutions of the dynamics equations (\ref{eq:nls2D_ESNL}) or Eq. (\ref{eqn:EulerBernoulli}--\ref{eqn:Eulercontinuite}). Note that stationary solutions are those of the equivalent diffusive equations that read in the abbreviated form
\begin{equation}
\frac{\partial \Psi}{\partial t} = \Ld \Psi + \Wd(\Psi)
\label{abbrevRelax}
\end{equation}
In the general case ($\xi\neq 0$), we have   
\begin{align}
&\Psi\equiv\psi\new \qquad \qquad \Ld\equiv\Delta\\
\begin{split}&{\Wd}(\psi\new)\equiv 
\left\{ -\frac{c}{\sqrt 2\xi} (|\psi_n|^2-1)\psi_n - \id\vvec\cdot\nabla\psi_n \right.\\ 
&\qquad\qquad\left.- \frac{c}{\sqrt{2} \xi}\left[ \xi^2 (\nabla\Phi_0)^2\psi_n + \xi^2 2\id (\nabla\Phi_0)\nabla\psi_n\right] + \vvec\cdot(\nabla\Phi_0)\psi_n\right\}\label{eq:termeNL_GLR_NLS}
\end{split}
\end{align}

In the Eulerian case ($\xi=0$), as discussed in section \ref{sec:euler_limit}, these definitions reduce to 
\begin{align}
&\Psi \equiv \phi \qquad\qquad\Ld \equiv \Delta
 \qquad\qquad\Wd \equiv  \nabla(\varrho\nabla\phi) - \vvec\cdot\nabla\phi\label{Wdefrelaxeuler}
\end{align}
with
\begin{align}
\varrho = -\frac 1 2 (\nabla\phi)^2  + \vvec\cdot \nabla\phi
\end{align}
To integrate (\ref{abbrevRelax}) a mixed implicit-explicit first-order 
time stepping scheme is used:
\begin{equation}
\label{time_step}
\Psi(t+\tau) = (\mathrm{I}-\tau \Ld)^{-1} (\mathrm{I}+\tau \Wd)\Psi(t)
\end{equation}
where $\Id$ is the identity operator and $\tau$ the time step.

The Helmholtz operator $(\Id-\tau \Ld)$, block-diagonal
with respect to Fourier modes, is easily
inverted in the Fourier--\cheb\ representation using the LU 
algorithm \cite{bib:NumRecNewtonRaphson}. 

As called for the $\tau$ method \cite{GO77}
the boundary conditions
(\ref{bc_Euler}),  (\ref{bc_NLSDirichlet}) or (\ref{bc_NLSNeumann}) are substituted to the equations 
 (\ref{abbrevRelax}) for the highest \cheb\ modes
$T_{N_r-1}$ and $T_{N_{r}}$.
The operator $(\Id-\tau \Ld)$ is thus modified before inversion.

This relaxation method 
can only reach stable stationary solutions of (\ref{abbrevRelax}).     
In order to also capture unstable stationary solutions \cite{Seydel}
we use the Newton branch-following method detailed in \cite{TB00}, \cite{HB97}, \cite{HB00}. 

\subsubsection{Branch following procedure}

We search for fixed points of (\ref{time_step}),
a condition strictly equivalent to the stationarity of (\ref{abbrevRelax}).
Each Newton step requires solving a linear system for 
the decrement $\psi$ to be subtracted from $\Psi$:

\begin{eqnarray}
\left[(\Id-\tau \Ld)^{-1} \right.&&\left.(\Id+\tau \Dd\Wd)-\Id\right]\psi \nonumber\\
&& = \left[ (\Id - \tau \Ld)^{-1} (\Id+\tau \Wd) - \Id \right]\Psi \label{Newton}
\end{eqnarray}
where $\Dd\Wd(\Psi)$ is the Fr\'echet derivative, or Jacobian matrix, of $\Wd$
evaluated at $\Psi$. Equation~(\ref{Newton}) is equivalent to:
\begin{equation}
(\Id-\tau \Ld)^{-1} \tau (\Ld+\Dd\Wd)\psi =
(\Id-\tau \Ld)^{-1} \tau (\Ld+\Wd)\Psi \label{Newton2}
\end{equation}
The role of $\tau$ is formally that of the time step
in (\ref{time_step}), but in (\ref{Newton}) or (\ref{Newton2}), 
its value can be taken to be arbitrarily large.
For $\tau \rightarrow \infty$, (\ref{Newton2}) becomes:
\begin{equation}
\Ld^{-1} (\Ld+\Dd\Wd)\psi =\Ld^{-1} (\Ld+\Wd)\Psi \label{Newtoninf}
\end{equation}

In order to solve the linearized systems stemming from the Newton method, we use BiCGSTAB \cite{Vandervorst}. 
We vary $\tau$
empirically to optimize the preconditioning and convergence of BiCGSTAB.
A few hundred BiCGSTAB iterations are usually
required to solve the linear system.

\subsubsection{Dynamics}
We write Eq. (\ref{eq:nls2D_ESNL}) in the abbreviated form:
\begin{equation}
\frac{\partial \Psi}{\partial t} = \Ld' \Psi + \Wd'(\Psi)
\label{abbrevDyn}
\end{equation}
where
\begin{align}
&\Psi \equiv \psi\new\qquad\qquad \Ld' \equiv  -\id \Ld\qquad\qquad\Wd' \equiv -\id \Wd(\psi\new) \label{Wdef}
\end{align}
with $\Wd$ defined in equation (\ref{eq:termeNL_GLR_NLS}) 

Equation (\ref{abbrevDyn}) is time stepped using the implicit Euler scheme
\begin{equation}
\Psi_{n+1} =(1- \tau \Ld)^{-1}\left[\Psi_{n} + \tau \Wd_n\right]
\label{LFCN}
\end{equation}
The boundary conditions
(\ref{bc_Euler}),  (\ref{bc_NLSDirichlet}) or (\ref{bc_NLSNeumann}) are imposed by modifying the operator
$(\Id-\tau \Ld)^{-1}$, as done for the relaxation 
time stepping algorithm (\ref{abbrevRelax})~\cite{GO77}.
\subsection{Numerical convergence}
\subsubsection{Euler}

Table \ref{tab:errMc} shows the error on our reference Mach number versus $N_\theta$ and $N_r$. Note that the errors are mainly due to a lack of Fourier modes in $\theta$. Thus, when a sufficient number of Fourier modes is reached, increasing the resolution in $r$ yields a better precision.

\begin{table}[ht!]
\begin{center}
\small{\begin{tabular}{|c|c|c|c|c|c|c|c|}
\cline{3-8}
\multicolumn{2}{c|}{}&\multicolumn{6}{c|}{$  N_\theta$}\\
\cline{3-8}
\multicolumn{2}{c|}{}&$  16$&$32$&$64$&$128$&$256$&$512$\\
\hline
      &$  16$& $4{.}45\times 10^{-3}$    & $3{.}72\times 10^{-4}$& $1{.}02\times 10^{-5}$ & $2{.}59\times 10^{-7}$ & $2{.}27\times 10^{-7}$ & $2{.}27\times 10^{-7}$ \\
\cline{2-8}
$N_r$       &$  24$& $4{.}45\times 10^{-3}$    & $3{.}72\times 10^{-4}$& $9{.}97\times 10^{-6}$  & $3{.}34\times 10^{-8}$ & $5{.}87\times 10^{-10}$ &$2{.}22\times 10^{-10}$  \\
\cline{2-8}
  &$  32$ & $4{.}45\times 10^{-3}$    & $3{.}72\times 10^{-4}$& $9{.}97\times 10^{-6}$ & $3{.}32\times 10^{-8}$ & $2{.}70\times 10^{-12}$ & $0$ \\
\hline
\end{tabular}}
\end{center}

\caption{Relative error versus resolution on the critical Mach number calculated by taking as a reference $\mathcal{M}\crit = 0{.}36969705259(9)$ calculated at ($512\times 32$). 
}
\label{tab:errMc}
\end{table}

Due to the use of \cheb\ polynomials, the boundary layers of the NLS flow computed at low Mach number are well resolved thanks to the large number of collocation points at the vicinity of the obstacle, as we will see in next section.

\subsubsection{Small coherence length solutions}\label{sec:app_small}

\begin{figure}[ht!]
\centerline{ \includegraphics[width=.45\textwidth]{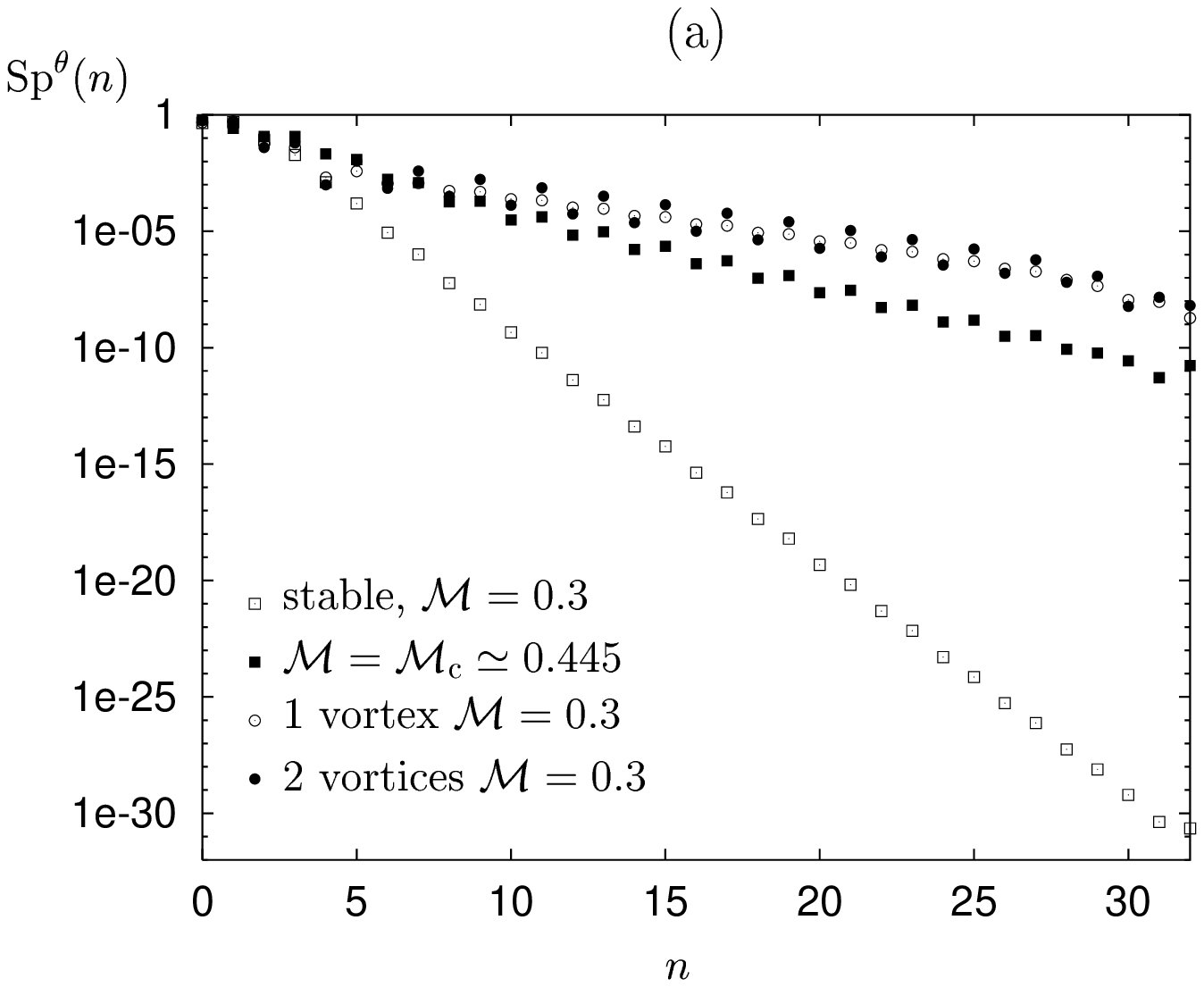}
\qquad  \includegraphics[width=.45\textwidth]{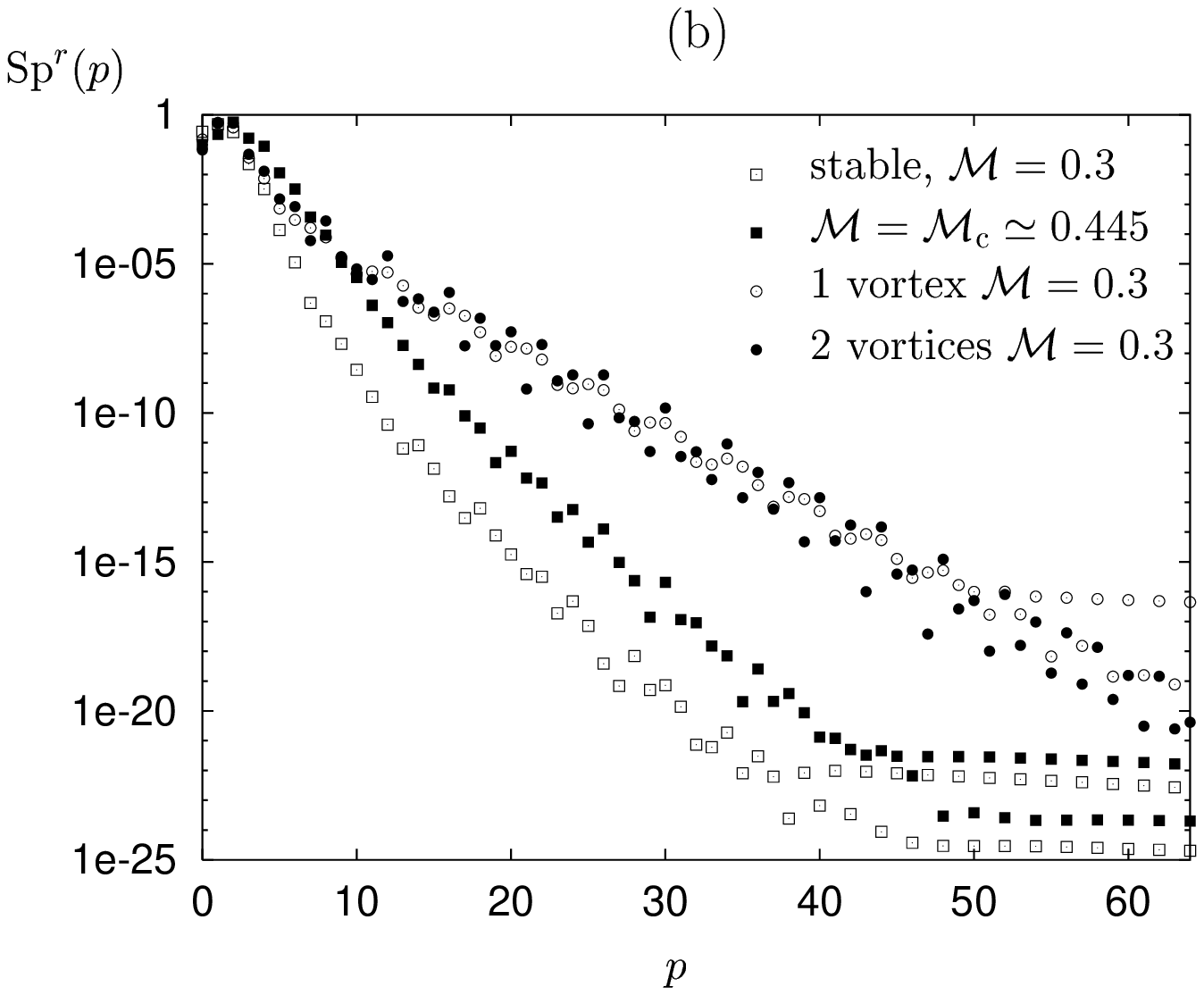}}
\caption{Stationary solution spectra with $\xi/D = 1/20$, $N_\theta\times N_r = 64\times 64$,
and Neumann boundary conditions: (a) $\theta$ spectra and (b) $r$
spectra for a stable solution far from the bifurcation,
for the solution at the bifurcation, and for a one vortex and a two vortex
unstable solutions. Spectral convergence is achieved for all stationary solutions.
}
\label{fig:nls_2D_spectres_small_xi}
\end{figure}

Our numerical method based on \cheb\ polynomial expansions allows to
solve the boundary layer of order $\xi$ by refining the collocation points
near the boundary conditions. The smaller $\xi$, the larger the radial resolution
$N_r$ must be. The azimuthal resolution $N_\theta$ depends also on the value of
$\xi$ through the multiplication of complex fields with a phase term such that
$\Phi_0(\theta,r) = v\frac{\cos\theta}{\sqrt 2 c\xi r}$ and $
\psi\new = \psi\old \ed^{\id\Phi_0}.$
 The phase term $\Phi_0$ is inversely proportional to $\xi$ and
needs sufficient $N_\theta$ points in order to be resolved.
Table~\ref{tab:nls_2D_CV} lists the resolutions used for computing the
bifurcation diagram for each $\xi/D$.
Spectral convergence is achieved for all stationary solutions as shown
in figure~\ref{fig:nls_2D_spectres_small_xi}.

\begin{table}[ht!]
\begin{center}
\begin{tabular}{c|| c c c c c}
$\xi/D$ & $1/2$ & $1/20$ & $1/40$ & $1/80$ & $1/120$\\
\hline
$N_\theta \times N_r$ & $ 64 \times 64 $ & $ 64 \times 64 $ & $ 128  \times 128 $ & $ 128 \times 128 $ & $ 256 \times 128 $ 
\end{tabular}
\end{center}
\caption{Azimuthal and radial resolutions used for computing the bifurcation diagram for
different $\xi/D$ for the two types of boundary conditions.}
\label{tab:nls_2D_CV}
\end{table}

\subsubsection{Large coherence length solutions}

For large ratio $\xi/D$, we have modified the mapping in order to stretch in the radial direction the collocation points next to the obstacle (see equation (\ref{mapgeneralise})). The choice of the value of $\lambda$ depends on the value of $\xi/D$ and the resolution $N_\theta \times N_r$ of the system.

\begin{figure}[ht!]
\centerline{ \includegraphics[width=0.45\textwidth]{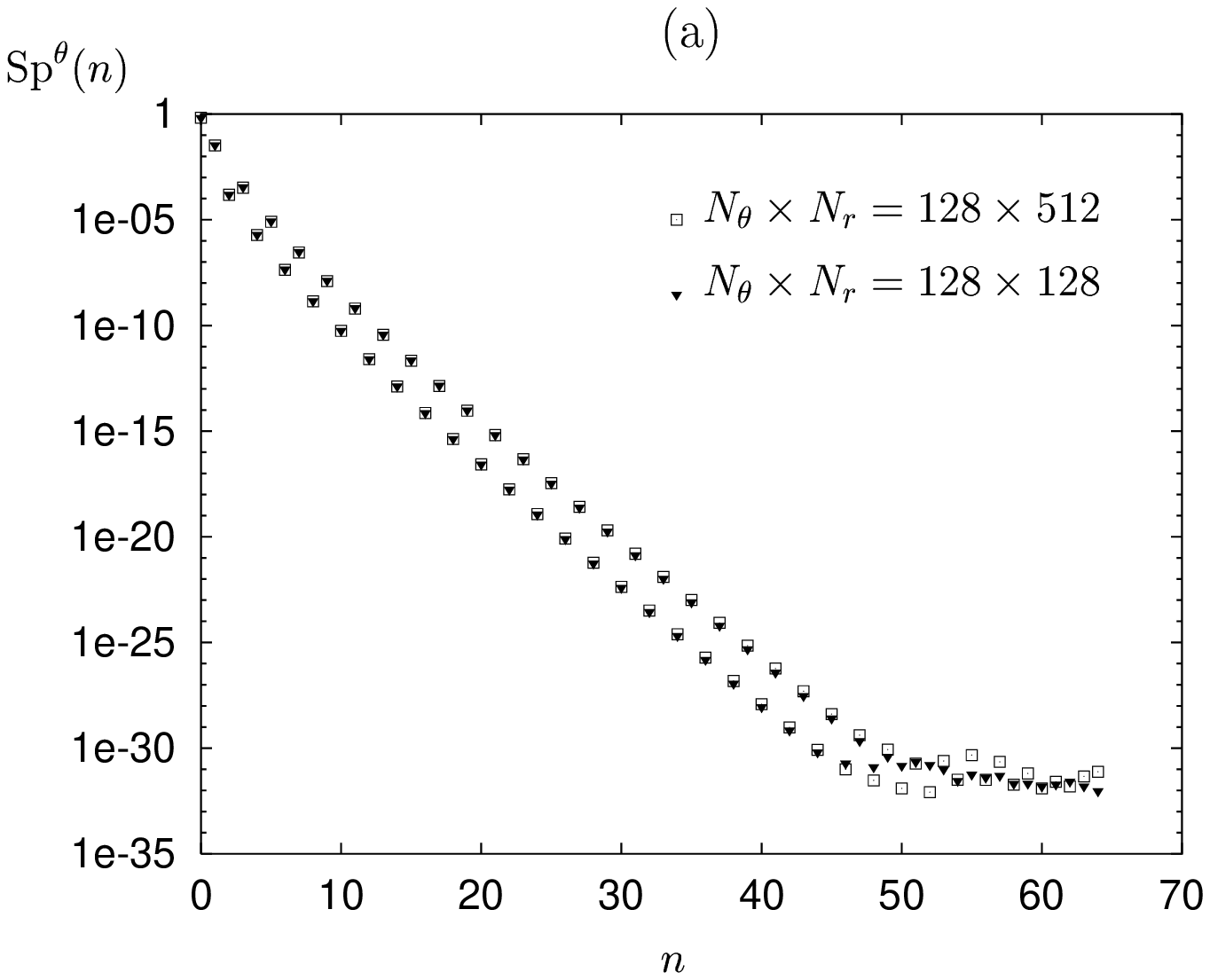}\hfill  \includegraphics[width=0.45\textwidth]{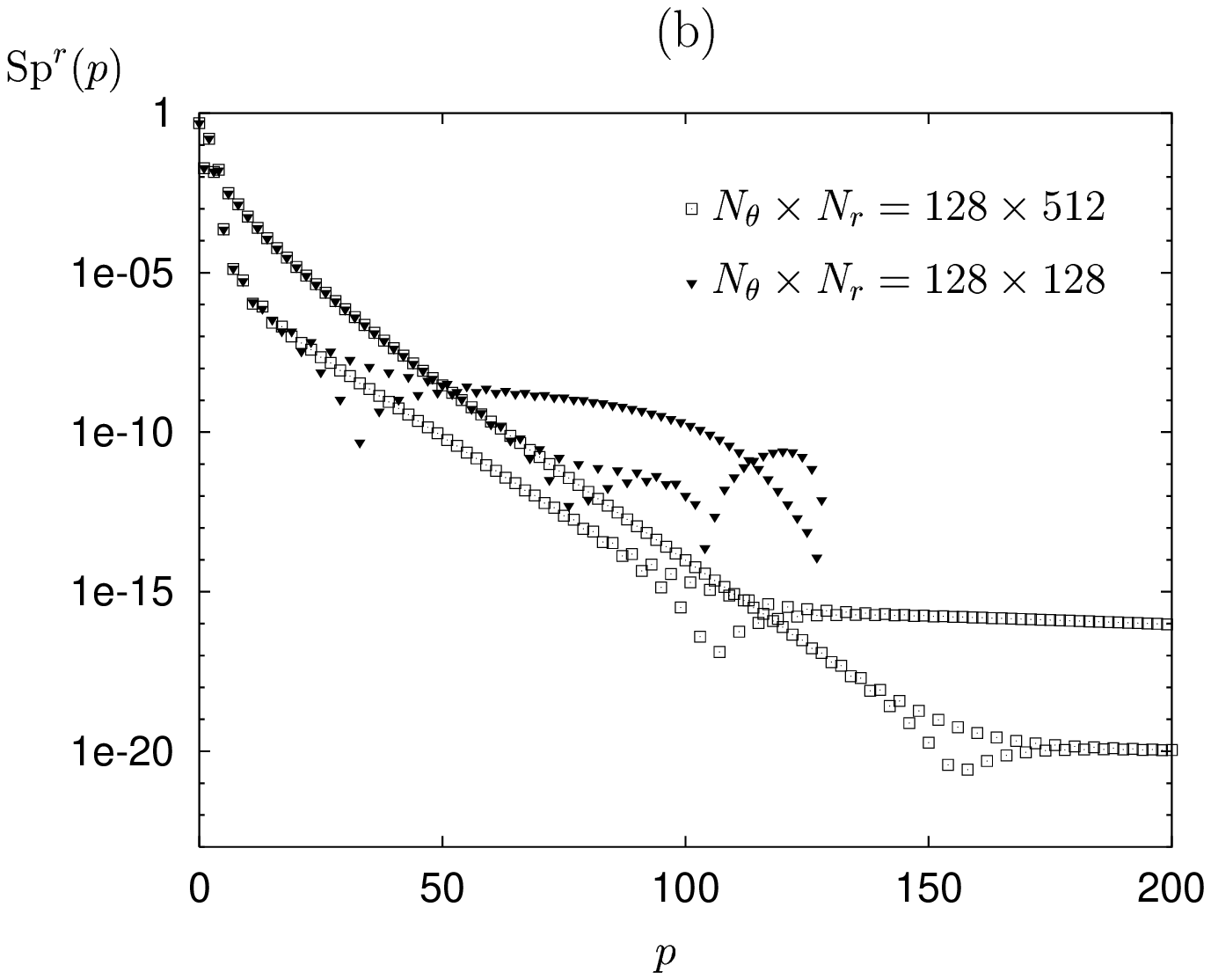}}
\caption{Symmetric unstable stationary solution spectra with $\xi/D = 20$, for two resolutions
($N_\theta\times N_r = 128\times 128$ et $128\times 512$)
and Dirichlet boundary conditions: (a) $\theta$ spectra and (b) $r$ spectra.
Spectral convergence is achieved for all stationary solutions. 
}
\label{fig:nls_2D_spectres_large_xi}
\end{figure}

Figure~\ref{fig:nls_2D_spectres_large_xi} shows azimuthal and radial spectra
for a symmetric unstable stationary solution at $\xi/D = 20$ with
a dilatation parameter $\lambda=80$ for two different resolutions.
Spectral convergence is achieved for all stationary solutions.

\bibliographystyle{unsrt}
\bibliography{biblisort,biblio}
\end{document}